\documentclass[draft,wrr]{agujournal2019}
\usepackage{amsmath}
\usepackage[english]{babel}
\usepackage{booktabs}
\usepackage{fontenc}
\usepackage[finalnew]{trackchanges} %for better track changes. finalnew option will compile document with changes incorporated. inline: with marked changes
\usepackage{lineno}
\usepackage{microtype}
\usepackage{ragged2e}
\usepackage{rotating}
\usepackage{soul}
\usepackage{upgreek}
\usepackage{url}
\draftfalse
\journalname{Water Resources Research}

\begin{document}
\addeditor{author}

\title{Multi-Tracer Groundwater Dating in Southern Oman using Bayesian Modelling}

\authors{Viola Rädle\affil{1,5}, Arne Kersting\affil{1}, Maximilian Schmidt\affil{1,2}, Lisa Ringena\affil{2}, Julian Robertz\affil{2}, Werner Aeschbach\affil{1}, Markus Oberthaler\affil{2}, Thomas Müller\affil{3,4}}

\affiliation{1}{Heidelberg University, Institute of Environmental Physics}
\affiliation{2}{Heidelberg University, Kirchhoff-Institute for Physics}
\affiliation{3}{Helmholtz Centre for Environmental Research, Leipzig}
\affiliation{4}{Helmholtz Centre for Ocean Research, GEOMAR, Kiel}
\affiliation{5}{Leipzig University of Applied Sciences (HTWK)}

\correspondingauthor{Viola Rädle}{viola.raedle@web.de}

\begin{keypoints}
\item Groundwater in a semi-arid area was dated with multiple tracers including the first full-scale application of 39Ar with Argon Trap Trace Analysis
\item Bayesian Inference was applied for modelling the transit time distributions using a Markov-Chain Monte Carlo simulation
\item A Dispersion Model with two components and a nonparametric model with six age bins were applied, both suggesting a mixed groundwater of very old and very young origin
\end{keypoints}

%\graphicspath{{C:/Users/Viola Rädle/Masterarbeit/Paper/Plots/}}
%\graphicspath{{../Plots/}}

\justify
\begin{abstract}
In the scope of assessing aquifer systems in areas where freshwater is scarce, estimation of transit times is a vital step to quantify the effect of groundwater abstraction. Transit time distributions of different shapes, mean residence times and contributions are used to represent the hydrogeological conditions in aquifer systems and are typically inferred from measured tracer concentrations by inverse modelling. 
In this study, a multi-tracer sampling campaign was conducted in the Salalah Plain in Southern Oman including CFCs, SF$_6$, $^{39}$Ar, $^{14}$C and $^{4}$He. Based on the data of three tracers, a two-component Dispersion Model (DMmix) and a nonparametric model with six age bins were assumed and evaluated using Bayesian statistics. In a Markov Chain Monte Carlo approach, the maximum likelihood parameter estimates and their uncertainties were determined. Model performance was assessed using Bayes Factor and Leave-One-Out cross validation. 
Both models suggest that the groundwater in the Salalah Plain is composed of a very young component below 30 years and a very old component beyond 1000 years, with the nonparametric model performing slightly better than the DMmix model. All wells except one exhibit reasonable goodness of fit. Our results support the relevance of Bayesian modelling in hydrology and the potential of nonparametric models for an adequate representation of aquifer dynamics.
\end{abstract}

\section{Introduction}
According to the IPCC special report on Climate Change and Land \cite{IPCC2019}, drylands (arid, semi-arid and dry sub-humid areas) currently cover about 46\,\% of the global land. Since the 1960s, dryland areas have expanded on average by about 1\,\% annually, due to desertification. In those regions, surface water is often rare and groundwater is the main source of freshwater supply for agriculture, industry, households and vegetation. 

Under the influence of climate change, investigating groundwater systems in drylands is important in two aspects. First, about 3 billion people live in dryland regions where groundwater is often the main water source \cite{VanderEsch2017}. 
Thus, groundwater management is most important and with that, a good understanding of groundwater dynamics, recharge and its response to water abstraction. Secondly, groundwater serves as a climate archive in which important information like infiltration temperatures and indicators of rainfall patterns are preserved. This makes groundwater analysis an important puzzle piece for paleoclimate reconstructions \cite{Seltzer2021,Varsanyi2011}.

The \change[author]{here presented study}{study presented here} investigates the groundwater system of the Salalah Plain in the South of Oman. While 82\,\% of the country is dominated by a hyper-arid sandy land desert \cite{Al-Ajmi2013}, the Salalah region is categorized as semi-arid \cite{Al-Ajmi2018}, mainly due to the annual Indian monsoon passing over the South of Oman. Still, water is scarce, as water consumption has drastically increased over the past decades driven by agricultural production \cite{Shammas2007}. This leads to a depletion of the local aquifers and saltwater intrusions from the ocean \cite{Askri2016a, Bear1999a, Shammas2007}. Furthermore, climate change driven alterations in rainfall patterns and sea level rise are likely to increase the stress on the existing freshwater reservoirs \cite{Ahmed2012, Al-Habsi2014, Al-Sarmi2017, Gunawardhana2016}. 
To analyse and potentially mitigate the impacts of the water consumption on the aquifers, a profound understanding of the groundwater system of the Salalah Plain is crucial.  

A very useful and well-proven tool to analyse groundwater dynamics is the concept of Transit Time Distributions (TTDs) \cite{Kirchner2001, Maloszewski1982, McGuire2006}\remove[author]{(Sprenger et al., 2019)}. It assumes that the investigated water body represents a mixture of water with different ages, hence, travelling times between groundwater recharge and abstraction. The age distribution reflects certain properties of the groundwater system, like the ratio between advective and diffusive flow, the mean flow velocity or the travel distance from the point of recharge. 

For groundwater age modelling, a specific type of TTD based on the hydrological setting is assumed and the free parameters of that TTD are constrained with tracer measurements. Typically, a variety of age tracers are applied, each covering a specific age range depending on their geochemical properties and atmospheric concentration. In this study, eight \add[author]{dating }tracers were used to investigate six monitoring and pumping wells. CFC-11, CFC-12, CFC-113 \cite{Oster1996}\change[author]{ and}{,} SF$_6$ \cite{Busenberg2000,Newman2010}\add[author]{and $^3$H }\cite{Clark1997_Trit} were used to identify waters that infiltrated within the last 60 years, $^{39}$Ar was applied to cover the dating range between 100 and 1000 years \cite{Loosli1983, Ritterbusch2014} and $^{14}$C \cite{Clark_1997, IAEA2013_C14, Vogel1968} and $^{4}$He \cite{Solomon2000} were used for groundwater that recharged on the timescale of millennia. $^{39}$Ar is of special importance in this suite of tracers as it uniquely covers the intermediate age range. \citeA{Ritterbusch2014} demonstrated that $^{39}$Ar analysis of groundwater samples by Atom Trap Trace Analysis (ATTA) \cite{Lu2014} is feasible. This study represents the first application of the $^{39}$Ar-ATTA or ArTTA technique for a complete set of groundwater samples and its integration in comprehensive age modelling of an aquifer system.

% This study represents the first full-scale application of the promising Atom Trap Trace Analysis (ATTA) method  to a groundwater study.
% In this study, the $^{39}$Ar-ATTA or ArTTA technique is used for the first time to analyse a complete set of groundwater samples and to include it in the comprehensive age modelling of an aquifer system.

To derive a TTD from the measured tracer concentrations, modelling software of different complexity is available \cite{Jurgens2012, Suckow2012}. The modelling approach applied in the framework of this study is based on Bayesian statistics\change[author]{ and}{, using the tracers CFC-11, $^{39}$Ar and $^{14}$C. It} incorporates the modelling of multiple water masses represented by analytical TTDs as well as shape-free age distributions which have been increasingly applied in the recent past \cite{Massoudieh2014, McCallum2014}. Furthermore, the model is capable of finding maximum likelihood estimates using a Markov-Chain Monte Carlo simulation and  includes a comprehensive uncertainty analysis. To compare the models' performance in terms of accuracy and simplicity, statistical key quantities to assess efficiency and predictive adequacy are computed. Evaluating the tracer data of the six wells with the described model allows for a \change{systematical}{systematic} analysis of possible age distributions and with that, a profound interpretation of probable groundwater dynamics in the Salalah Plain.

\begin{figure*}
    \centering
    \includegraphics[width=0.85\linewidth]{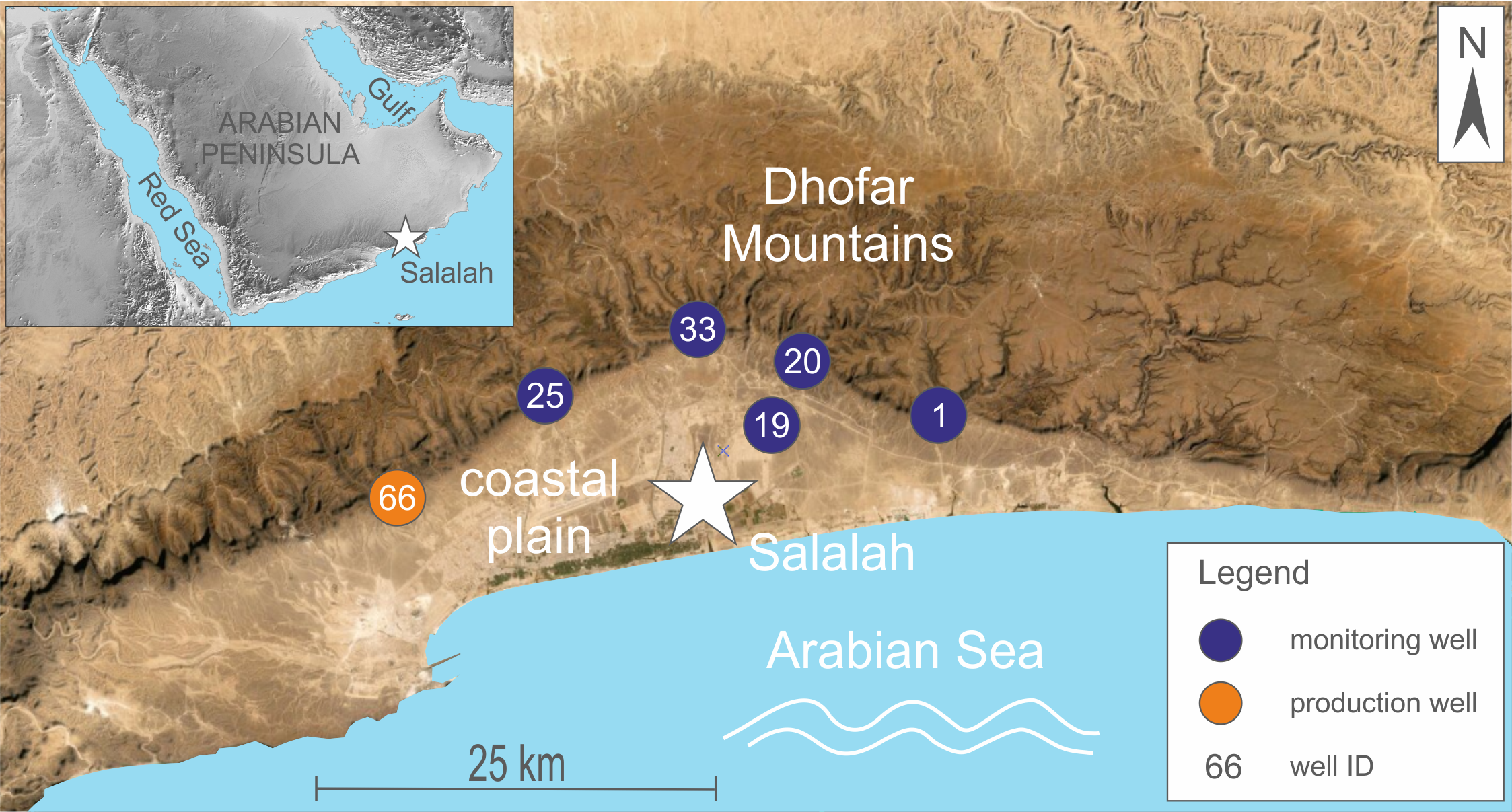}
    \caption{Sampled wells in the Salalah Plain on the foot of the Dhofar Mountains.}
    \label{fig:sampled_wells}
\end{figure*}

\section{Methods}

\subsection{Site description and sample processing}

The Salalah Plain in Southern Oman forms a crescent shaped coastal enclave, with the Dhofar Mountain Range in the North and the Arabian Sea in the South. Being 60\,km long and 15\,km wide at its broadest point, the throughout flat plain is host of the City of Salalah with \change[author]{over}{its around} 300\,000 \change[author]{people the second largest city in the Sultanate of Oman}{inhabitants}. The Dhofar Mountains are about 15\,km wide in North-South direction. They rise steeply from the plain to elevations of around 900 meters in the center, separating the semi-arid south from the arid north. 
The area is affected by the Indian Summer Monsoon, with the mountain ridge as the Northern limit for the moisture. 
Precipitation is low on the plains ($<\,$100\,mm) and high in the mountains (up to 450\,mm), with almost all precipitation during the monsoon between June and September \cite{Shammas2007}. The monsoon represents the most reliable water source of the study area \cite{Bawain2012, Clark_1987, DandMI1992}. Additionally, occasional storm events can bring significantly more rainfall in a few days than in the rest of the year \cite{Friesen2018, Muller2020}. Those storms can cause vast flooding, leading to direct groundwater recharge in the plain.
Karstified limestones form the aquifers in the mountains and in the coastal plain. The mountains are the main recharge area and groundwater reservoir, where the southward flow is higher in the center than in the East and West \cite{GeoResourcesConsultancy2004}. The aquifer underlying the plain is the Fars formation, a shallow karstic limestone aquifer of tertiary origin with a saturated thickness of up to 70 meters. Groundwater users are the public water supply, agriculture and industry, with the main abstraction being in the central part of the plain, resulting in a deficit in the coastal water balance \cite{DandMI1992, ICBA}.

In the study area, six groundwater \change[author]{samples were taken, five from monitoring wells and one from a pumping well}{wells were sampled, five monitoring wells and one pumping well} (Figure \ref{fig:sampled_wells}). 
Before sampling the wells, the stagnant water was removed using a \textit{Grundfos MP1} submersible pump to prevent atmosphere-equilibrated water from contaminating the samples. \add[author]{\protect Temperature, pH, electrical conductivity (EC) and dissolved oxygen (Multi 3430, WTW, Weilheim, Germany) were measured on-site. The alkalinity was determined by titration with H$_2$SO$_4$ (AL-DT alkalinity test kit by HACH, Düsseldorf, Germany) and converted into bicarbonate concentrations. The wells were sampled for the analysis of ion chemistry, water stable isotopes, noble gases, carbon isotopes, tritium, $^{39}$Ar and transient trace gases. As this study focuses on groundwater dating, the geochemical data (water chemistry, stable isotopes, atmospheric noble gases) are not further discussed, except for their use in the interpretation of the tracers that can provide groundwater age information, i.e. radiogenic $^4$He, $^{14}$C, $^{39}$Ar, $^{3}$H, CFCs and SF$_6$. For each well and dating tracer, duplicate samples were taken.}

\remove[author]{The wells were sampled for analysis of $^4$He, $^{13}$C and $^{14}$C, $^{39}$Ar, $^{3}$H, CFCs and SF$_6$. In addition, the temperature, pH, electric conductivity, titration and ion composition were measured to investigate the groundwater chemistry. For each well and tracer, duplicate samples were taken.}
\remove[author]{The measured $^3$H concentrations were hardly significantly different from zero, therefore did not allow reliable differentiation between the wells (Table DS01). For the study region, it is very difficult to construct a reliable tritium input function for the entire relevant period including the bomb peak period, as no GNIP stations are nearby and even those from far away around the Indian ocean have only incomplete data \protect \cite{IAEA/WMO2021}.} \remove[author]{For those reasons, tritium measurements were excluded from analysis.}

\textbf{Radiogenic \textsuperscript{4}He} For noble gas sampling, crimped copper tubes, pinched off with stainless steel clamps for vacuum tight sealing, were used. The concentrations of stable noble gas isotopes were measured using mass spectrometry at the Institute of Environmental Physics, Heidelberg University. Using the software \textit{PANGA} \cite{Jung2018}, noble gas temperatures and helium components were modelled. 

\textbf{\textsuperscript{13}C and \textsuperscript{14}C} The carbon samples were taken using evacuated 100\,mL glass bottles with a needle pinched through a rubber sealing. To avoid biological activity in the water sample, silver nitrate was added. At Heidelberg University, the carbon dioxide was extracted and reduced to carbon by graphitization. At the Curt Engelhorn Centre for Archeometry Mannheim, the \textsuperscript{13}C and \textsuperscript{14}C concentrations were measured using accelerator mass spectrometry (AMS). 

\textbf{\textsuperscript{39}Ar} For \textsuperscript{39}Ar sampling, an evacuated 12.2\,L steel bottle was chosen as a suitable container, with a three-way brass adapter for flushing \cite{Beyersdorfer2016, EbserPaper, Radle2019}. At Heidelberg University, the gas was extracted, the reactive gases were removed by gettering and the noble gases (about 2\,mL) were stored in an Activated Charcoal Trap (ACT). The \textsuperscript{39}Ar concentration was measured using Argon Trap Trace Analysis (ArTTA), exploiting the high isotopic selectivity of the atom optical technique \cite{Chen1999, Feng2019, Lu2014, Ritterbusch2014}. Single \textsuperscript{39}Ar atoms are captured in a magneto-optical trap (MOT) and are identified by the emission of fluorescence photons. The count rate of the sample is compared to a reference, giving the concentration in percent modern argon (pmAr). 

\add[author]{\textbf{$^3$H:} The samples were analyzed for tritium by means of Liquid Scintillation Counting after electrolytic enrichment. Concentrations were reported in Tritium Units (TU).}

\textbf{CFCs and SF\textsubscript{6}}
The samples for Chlorofluorocarbons (CFCs) and Sulfur hexafluoride (SF\textsubscript{6}) analysis were stored in 500\,mL stainless steel cylinders with ball valves. The concentrations were derived using gas chromatography (GC). Duplicate samples were taken of which the A samples were measured at Spurenstofflabor Wachenheim, Germany while the B samples were measured at Heidelberg University. 

\subsection{Transit time distributions}

To characterize the dynamics of an aquifer system, dating tracers are applied to gather information on the time scales on which the subsurface flow takes place. The idealized \textit{groundwater age} is defined as the time elapsed between the entering of water into the saturated zone and its sampling at a distance downstream within the groundwater system \cite{IAEA2013}. However, one distinct value for the age (\textit{tracer age}) only describes a purely advective regime with a point like recharge area. In reality, longitudinal and transverse dispersion leads to mixing along the flow path and between different flow paths, yielding a complex age spread denoted as transit time distribution (TTD) \cite{Hall1994}.

The simplest TTD is the Piston-flow Model, where any kind of mixing is neglected (pure advection) and the screen of the well is infinitesimally small. Mathematically, this TTD is represented by a \add[author]{Dirac }delta distribution \change[author]{with a probability of 1 for}{around} the mean age.
In contrast, the Exponential Model is used for unconfined aquifers with a screen expanding from the bottom to the top \cite{Vogel1968}. All flow lines are considered, with \change[author]{a weight of the input events}{the contribution of recharge} decreasing exponentially with age. 
The Dispersion Model is a solution for the advective-diffusive groundwater transport equation \cite{Maloszewski2000}. The two  parameters are the mean residence time $\tau$ and the Péclet number $Pe$, a measure of the relative importance of advection compared to the dispersion within a system \cite{Waugh2003}. As the Péclet number works as a shape parameter, the Dispersion Model approaches the Exponential Model for low \textit{Pe} and the Piston-flow Model for large \textit{Pe}. 
\add[author]{Combinations of the above described TTD models or truncations of their age distributions are sometimes used for specific hydrogeologic situations \protect \cite{Jurgens2012}.} In addition to analytical age distributions\add[author]{ for a single groundwater component}, mixing of several distinct water components from different reservoirs can occur. This is adopted by \add[author]{binary mixing models combining two of the above models and }weighting the individual TTDs by \change[author]{their}{the} mass contribution\add[author]{ of the components \protect \cite{Jurgens2012}. Given that the dispersion model can describe flow conditions between no mixing (Piston-flow) and complete mixing (Exponential Model), we consider a binary mixing model that combines two independent Dispersion Models to be a flexible approach to approximate many realistic cases of groundwater systems.} 

If a water mixture cannot be described by analytical models, complex numerical methods can offer a broader range of possibilities \cite{Broers2021, Cirpka2007, Troldborg2008, Visser2013}. Those shape-free models, also called nonparametric models or age histograms, can help to identify the age contributions from a selected number of bins. 

\subsection{Multi-tracer modelling approach}

\subsubsection{Bayesian Inference and Maximum Likelihood}
To address uncertainty in hydrological modelling, Bayesian methods have been widely applied \remove[author]{in literature }\cite{Bates2001, Engeland2002, Liu2007, Vrugt2003}.
In Bayesian statistics, the probability \add[author]{$P$ }of a model $M(\theta)$ given the data $c$, also called the posterior, is computed by 
\begin{equation}
     P(M(\theta)\mid c) = {\frac{\mathcal{L}(c\mid M(\theta))\; P(M(\theta))}{P(c)}} 
\end{equation}
with evidence $P(c)$, prior $P(M(\theta))$ and likelihood $\mathcal{L}(c \mid M(\theta))$ \cite{Clark2005} whereas $M(\theta)$ denotes the shape of the TTD as well as an attendant set of model parameters $\theta$. The data $c$ are the measured tracer concentrations with measurement uncertainty.

As the exact form of the age distribution is not known, it can only be approximated by different presumed distribution functions that can be compared to one another. The \textbf{prior} $P(M(\theta))$ contains the presumed distribution function and the parameter space $\Theta$ given the constraints. In the present case, the mathematical models are a nonparametric model with six age bins and the mixture of two Dispersion Models with varying Péclet numbers\add[author]{ and mean ages} (see Section \ref{meth:math_model}). As no hydrological information is available before modelling, the prior is assumed uniform within the parameter constraints, resulting in a posterior proportional to the likelihood.
The \textbf{likelihood} $\mathcal{L}(c \mid M(\theta))$ denotes the probability of observing a tracer configuration $c$ conditioned to a transit time distribution $M$ and a set of parameters $\theta$. It is not normalized, and for numerical reasons the logarithm (log-likelihood) is used \cite{Kolanoski2002}: 

\begin{equation}
    \ln{\mathcal{L}(c \mid \theta)} =
    \ln{\mathcal{L}(c_\mathrm{1},...,c_\mathrm{n} \mid \theta)} = \sum_\mathrm{j=1}^\mathrm{n} \ln{l(c_\mathrm{j} \mid \theta)}
\end{equation}   

with number of tracers n, $c_\mathrm{j}$ representing the measured concentration of tracer j, and  $l(c_\mathrm{j} \mid \theta)$ the tracer likelihood of measurement $c_\mathrm{j}$ given the parameter set $\theta$. Note that this is a modified realization to the standard definition found in literature as n does not denote the number of data points. Instead, each model evaluation is based on one data point in n-dimensional tracer space. 

To calculate $l(c_\mathrm{j} \mid \theta)$, the error of $c_\mathrm{j}$ can be expressed in the form of a probability density \cite{Yustres2012}. The measurement uncertainty was used for all tracers except for \textsuperscript{14}C, where the uncertainty was enhanced due to the applied Vogel model with initial concentration of $80 \pm 5$\,\% \cite{Vogel1968}. The probability density was assumed to be normally distributed for all tracers except for \textsuperscript{39}Ar \change[author]{which follows Poisson/Gaussian statistics}{for which a dedicated Bayesian error analysis accounting for the Poissonian atom counting statistics as well as other sources of error was conducted} \cite{Ebser2018, Feng2019}. Hence, in lack of a parametrizable distribution, the numeric probability density distribution from the ArTTA measurements was used for \textsuperscript{39}Ar. The log-likelihood of tracer j is then determined by allocating $c_\mathrm{calc,j}(\theta)$, calculated by the mathematical model, in the probability density of the measurement data.

To calculate the \textbf{evidence} $P(c)$, a scalar normalization constant representing the probability of the data, it is necessary to integrate over the entire parameter space. However, the integral cannot always be evaluated, a problem known as intractable normalization. To approach this issue, the integral was approximated by summing over all parameter configurations.

An according approach is posed by the \textbf{maximum likelihood estimate (MLE)}, a quantity independent of the evidence as the normalization does not change throughout the analysis \cite{Cappe2002, Geyer1992}. The corresponding parameter estimates $\hat{\theta}$ represent the realization of a given model that comes closest to the measurements (conditioned to the limits of the parameter space). 

As an additional, more intuitive quantity for the agreement of the calculated and the measured tracer concentrations, a \textbf{likelihood score} is calculated. Given as the likelihood of the estimates $\mathcal{L}(c \mid \hat{\theta})$ divided by the (model independent) likelihood of the exact measured tracer data $\mathcal{L}_\mathrm{meas}$, it serves as a relative measure and will be expressed in \% throughout this paper.

\subsubsection{Mathematical model} \label{meth:math_model}
A mathematical model is deployed to calculate the tracer concentrations given the shape of the TTD and a parameter set $\theta$. First, for the tracers involved in the modelling approach the concentration in water $c_\mathrm{wa, j}(t_\mathrm{i})$ is derived for all (advective) ages $t_\mathrm{i}$, taking into account solubility and exponential decay. The solubility of CFC-11 in water is calculated according to \citeA{Warner1985}, using the Northern Hemispheric atmospheric input functions \cite{NOAA2018} and \change[author]{the measured}{modelled} input parameters. For $^{39}$Ar and $^{14}$C, radioactive decay is considered while taking into account a correction model for $^{14}$C (i.e. the Vogel model). Both are measured in relation to modern concentrations, with high values reflecting young water. 

The two TTDs employed in the modelling approach are a Dispersion Model with two water masses (DMmix) and a nonparametric model. 
In a Dispersion Model, the age distribution function follows an inverse Gaussian regime:
\begin{equation}
    g(t_\mathrm{i}, \tau, Pe) = \sqrt{\frac{Pe \cdot \tau}{4 \pi t_\mathrm{i}^3}} \cdot e^{-\frac{Pe \cdot (t_\mathrm{i} - \tau)^2}{4 \tau t_\mathrm{i}}}
\end{equation}
with time $t_\mathrm{i}$ in years before present, mean residence time (MRT) $\tau$ and Péclet number $Pe$. Mixing two water masses with different $\tau$ and $Pe$ gives the (normalized) age distribution
\begin{equation}
    \rho(t_\mathrm{i}) = r \cdot g(t_\mathrm{i}, \tau_1, Pe_1) + (1-r) \cdot g(t_\mathrm{i}, \tau_2, Pe_2)
\end{equation}
with $r$ denoting the mixing ratio of the first water mass, yielding a parameter vector $\theta = (\tau_1, Pe_1, \tau_2, Pe_2, r)$. The calculated concentration of tracer j is derived by
\begin{equation}
    c_\mathrm{calc, j}(\theta) = \sum_{t_\mathrm{i}=0}^{t_\mathrm{max}} \rho(t_\mathrm{i}) \cdot c_\mathrm{wa, j} (t_\mathrm{i})
\end{equation}
In case of presumed Péclet numbers, the parameter vector is reduced to $\theta = (\tau_1, \tau_2, r)$, leaving three free parameters for the DMmix model. \\

For the nonparametric model, the concentration of tracer j (in water) in bin k (normalized by bin size) is computed by
\begin{equation}
    \rho_\mathrm{k, j} = \frac{\sum_{t_\mathrm{i}=T_\mathrm{k-1}}^{T_\mathrm{k}}  c_\mathrm{wa, j} (t_\mathrm{i})}{T_\mathrm{k} - T_\mathrm{k-1}}
\end{equation}
for $t_\mathrm{i}$ between the age limits \{$T_\mathrm{k-1}$ and $T_\mathrm{k}$\} . The age limits can be chosen as required, depending on the desired resolution in the respective age regimes. 
With this, the concentration of tracer j is calculated:
\begin{equation}
    c_\mathrm{calc, j} = \sum_\mathrm{k=1}^\mathrm{K} \rho_\mathrm{k, j} \cdot \theta_\mathrm{k}
\end{equation}
with $\theta_\mathrm{k}$ being the fraction of groundwater in bin k. Hence, for a nonparametric model with K = 6 bins\add[author]{ and given bin boundaries}, there are five free parameters $\theta = (\theta_1, \theta_2, ... \theta_5)$ with fraction in the sixth bin $\theta_6 = 1 - \sum_\mathrm{k} \theta_\mathrm{k}$.

\subsubsection{Markov-Chain Monte Carlo methods}

Finding the parameters of a TTD that best describes the groundwater system and can reproduce the observed tracer data is a typical inverse problem. 
This is addressed by modelling possible tracer results with different TTDs and parameter sets to project the (non-analytical) likelihood function. 
The Maximum Likelihood Estimate, meaning the best parameter set $\hat{\theta}$, is found by minimization of the weighted distance between the observed and the simulated values, leading to an optimization problem \cite{Yustres2012}. Yet, the presence of multiple local optima may interfere with simple optimization methods. A solution to this is provided by Markov-Chain Monte Carlo (MCMC) methods, \change[author]{a Bayesian}{an} approach for parameter estimation and uncertainty analysis \cite{Zheng2016}. As they can sample from virtually any posterior distribution without the need for an analytical expression, MCMC methods have been applied in multiple hydrological studies \add[author]{\protect\cite{Kuczera1998, Laloy2013, Marshall2004, Massoudieh2012, Parno2014, Smith2008, Vrugt2008, Zheng2016}}. In a Monte Carlo simulation, a multitude of parameter samples are generated based on the posterior distribution \cite{Gamerman2006}. To advance the parameter samples towards the optimum, a Markov chain is applied, meaning that the acceptance probability of a current state only depends on the immediately preceding state, not on the entire path \cite{Sorensen2002}. The procedure of choice is the broadly applicable Metropolis Hastings algorithm, generating new parameter sets $\theta_\mathrm{m+1}$ depending on current one $\theta_\mathrm{m}$ (Markov Chain) \cite{Metropolis1953}, with a steady, normally distributed proposal density of width $\sigma_\mathrm{proposal}$. First, a sample $\theta'$ from the proposal density is drawn. In the next step, the likelihood of the proposed parameters $\mathcal{L}(c \mid \theta')$ is evaluated and compared to the previous one $\mathcal{L}(c \mid \theta_\mathrm{m})$. Depending on the acceptance probability $\alpha(\theta', \theta_\mathrm{m})$, the new parameters can be accepted (making $\theta'$ the new original state, $\theta_\mathrm{m+1} = \theta'$) or rejected (starting from the same point again, $\theta_\mathrm{m+1} = \theta_\mathrm{m}$). Hence, the proposal density controls both the step size and the acceptance rate. To reach the target of the Metropolis-Hastings algorithm i.e. the maximum, it is desired to have a large step size with a simultaneous high acceptance probability \cite{Gelman2013, Robert2004}.
As the computational cost of MCMC is relatively high, it is necessary to adjust the algorithm by modifying the proposal distribution and the acceptance probability to improve the search efficiency. A two-step procedure was applied with an initial broad scan of the parameter space (large step size, tolerant acceptance probability) and a tuning around the maximum found in the first step (small step size, strict acceptance probability). \add[author]{The MCMC algorithm was tested and tuned on synthetic data sets, as described in the Supporting Information (Text and Table S3).}

\subsubsection{Uncertainty of the fit results}
To review the parameter estimates $\hat{\theta}$, it is crucial to infer their probability density functions. Using maximum likelihood, the  uncertainty of the parameter estimates can only be calculated in particular cases. Generally, the whole covariance matrix is required \cite{Kolanoski2002}. For likelihood functions of known analytical forms, a Taylor series can be expanded around the maximum. As in this study, the likelihood does not correspond to known analytical distributions (i.e. the normal distribution), numerical methods are applied. To describe the likelihood regime, the moments of the function (mean, variance, skewness and kurtuosis) can be calculated. However, for a skewed distribution, the point of maximum likelihood does not correspond to the mean and due to constraints, the parameter space does not extend symmetrically to both sides. Therefore, this method is disregarded. Hence, the method of choice is the analysis of likelihood contours in parameter space. Those are hypersurfaces of constant likelihood forming an m-dimensional ellipsoid \cite{Kolanoski2002}. A confidence level of 68.27\,\% (corresponding to one standard deviation), is chosen as the partial integral of the likelihood within the error interval. The uncertainty of the parameter estimates is therefore derived as distance to the intersection of the parameter axes with the hypersurface.

\subsubsection{Model selection techniques} 

In the context of model comparison, a vital step is the assessment of the model performance. Two common techniques are applied in the scope of this study, the \textbf{Bayes factor} and \textbf{leave-one-out  cross validation}. \\
The Bayes factor corresponds to the goodness-of-fit across the entire parameter space and serves as a measure for the model performance \cite{Jeffrey1961, Kass1995}. Originally designed to compare the integrals of the likelihood given the prior density of two models, the Bayes Factor

    \begin{equation}
        B = \int_\mathrm{\Theta} \mathcal{L} (c \mid \theta) p(\theta) d\theta
    \end{equation}

will be expressed as the logarithm $\ln{\left(B \right)}$ in this paper. Computation of Bayes factors has been found to be difficult \cite{Ouarda2011}, not least due to the required time investment. Incidentally, the averaging process in $\ln{\left(B_\mathrm{k} \right)}$ implicitly penalizes complexity as a vast parameter space will generally contain large stretches of poor fit quality \cite{Vandekerckhove2015}. \\

\textbf{Leave-one-out (LOO) cross validation} is a cost intensive technique applicable for small data sets that is used to evaluate the predictive performance of a model more explicitly \cite{Gelman2014, Gronau2019, Stone1974}. 
To validate how a model outcome generalizes to independent data, the observations are repeatedly partitioned into a training set and a test set. Consecutively, one observation i is excluded from modelling and the fit is carried out with the remaining data only. The ascertained posterior is employed for observation i as test data to assess predictive adequacy. In this study, each well corresponds to one data point in n-dimensional tracer space. Hence, instead of leaving out one of several data points, it is necessary to leave out one measurement $c_\mathrm{i}$. 
The resulting LOO posterior $p(\theta\mid c_\mathrm{-i})$ is computed using all tracers except for tracer i, denoted as $c_\mathrm{-i}$, with the LOO log-likelihood 
\begin{equation}
    \ln{\mathcal{L}(c_\mathrm{-i} \mid \theta)} =
   \sum_\mathrm{j=1, \, j \neq i}^\mathrm{n} \ln{ l(c_\mathrm{j} \mid \theta)}
\end{equation}

The total expected log likelihood for leave-one-out cross validation $\ln \hat{\mathcal{L}}_\mathrm{LOO}$ is deduced by the sum over the individual tracers i \cite{Gelman2014, Gronau2019}: 
\begin{equation}
      \ln{\hat{\mathcal{L}}_\mathrm{LOO}} =  \sum_{\mathrm{i}=1}^3 \ln{\int_\Theta \mathcal{L}(c_\mathrm{i} \mid \theta) \, p(\theta \mid c_\mathrm{-i}) \, d\theta} 
\end{equation}
with $p(\theta\mid c_\mathrm{-i})$ being the LOO posterior computed with tracer data $c_\mathrm{-i}$ and $\mathcal{L}(c_\mathrm{i} \mid \theta)$ the likelihood of tracer $c_\mathrm{i}$. 
Bayesian computing includes sampling from the posterior over the whole parameter space, making the calculation very cost intensive. As a mesh grid, parameter step sizes of 1\,y, 1000\,y and 0.05 were chosen for MRT$_1$, MRT$_2$ and the ratio of the water masses, respectively. 
In practice, LOO-CV results can also be utilized to reveal the impact of the individual tracers on the parameter estimates, for a better understanding of the model. 

In general, poor prediction capability may be caused by overparameterized models, a phenomenon denoted as overfitting, resulting in a trade-off between the prediction capability and model complexity. Quantities like the Akaike and the Bayesian Information Criterion punish complexity by adding a term including the number of free parameters to the likelihood function \cite{Akaike1973, Schwarz1978}. Furthermore, the Deviance Information Criterion \cite{Draper1995} is frequently applied in hydrological modelling approaches \cite{Ju2020, Massoudieh2014, RezaNajafi2013}. However, several prerequisites for those criteria, e.g. a normal distributed likelihood in parameter space \cite{Spiegelhalter2002}, are not met by the here presented data, due to the mere fact that constraints set asymmetric boundaries.

\section{Results}

A summary of all measured tracer concentrations and field parameters is displayed in Table DS01 (Supporting Information). \add[author]{\protect The EC values range from 696 to 2270\,$\upmu$S/cm. Wells 20, 25 and 33 at the foot of the Dhofar Mountains display fresh groundwater while more saline water is found in the East (well 1), West (well 66) and in the lower central plain (well 19).
Regional salinity differences have been described since the 1970s \cite{Askri2016a, Macumber1998, Shammas2007} and led to the plain being divided into three zones, the 'Eastern Brackish Zone', the 'Central Freshwater Zone' and the 'Western Brackish Zone'. The origin of the salinization is not clear. It is assumed that, depending on the location and depth, different causes are responsible. Salt-bearing clays and marls occurring in the subsurface or the inflow of saline groundwater from deep aquifers triggered by reduced fresh water pressure in the pumped shallow aquifer come into question, but an oceanic origin (salt water intrusion due to groundwater extraction or previous flooding during times of higher sea-level) is also possible in some areas.  On the other hand, the low EC values of wells 20, 25 and 33 at the foot of the mountains are easy to explain. They are located in the main inflow area from the mountains into the plain, where throughflow from the mountains is highest \cite{Friesen2018}.} \\
The measured \add[author]{dating }tracer concentrations are systematically evaluated by first examining their individual informative value. Subsequently, \change[author]{the most robust set of tracers}{a subset of reliable tracers, as explained below,} is used for modelling a Transit Time Distribution and parameters that represent the flow in the Salalah groundwater system best.

\begin{figure*}
    \centering
    \includegraphics[width=\linewidth]{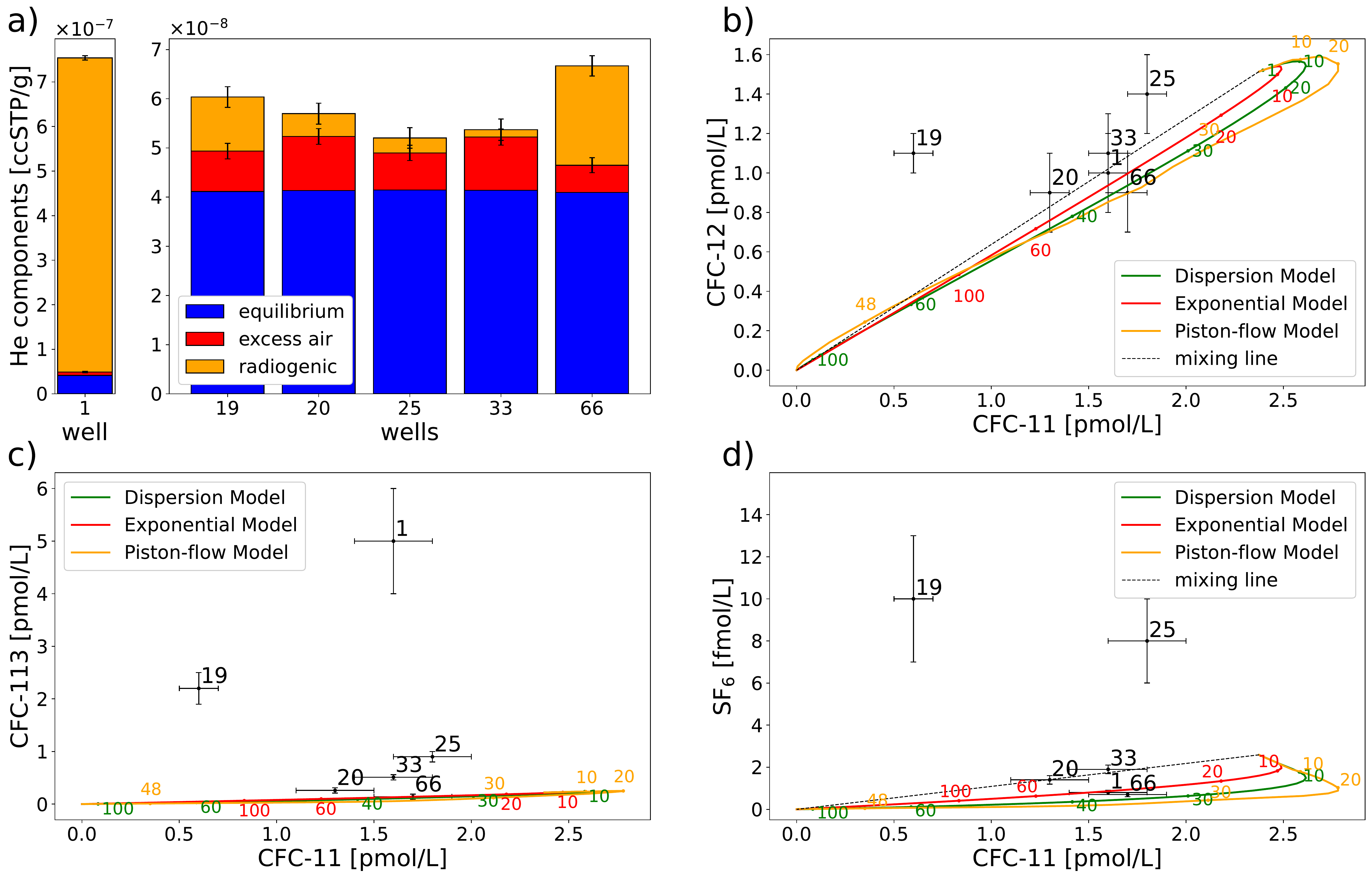}
    \caption{a) \textsuperscript{4}He components modeled with \textit{PANGA} using noble gas isotope measurements. Well 1 exhibits remarkably higher  $^{4}$He$_\mathrm{rad}$ levels than the other wells. b) Synoptic plot of CFC-11 and CFC-12 data and TTD trajectories with high\change[author]{accordance}{consistency} except for wells 19 and 25. c) Synoptic plot of CFC-11 and CFC-113 data with CFC-113 contamination in all wells except for well 66.  d) Synoptic plot of CFC-11 and SF\textsubscript{6} data with wells 19 and 25 above atmospheric SF$_6$ levels.}
    \label{fig_2}
\end{figure*}

\subsection{Tracers not involved in the modelling approach}

While the groundwater samples were tested for a variety of\add[author]{ dating} tracers, not all of them are suitable for a quantitative analysis. In Figure \ref{fig_2}, unemployable tracers can be identified and excluded from analysis.

\textbf{Radiogenic helium:} The bar chart in Figure \ref{fig_2} a) shows the components of \textsuperscript{4}He as modelled with the software \textit{PANGA} \cite{Jung2018}\change[author]{ using input parameters as described by Raedle (2019).}{\protect . The standard (CE model) noble gas data evaluation yields moderate excess air contributions for an  estimated infiltration altitude of 500\,masl. The assumed conditions (altitude) and those derived from the noble gases (temperature, excess air) are used as constraints for the calculation of radiogenic $^4$He and equilibrium concentrations of CFCs and SF$_6$. Otherwise, the stable noble gases only provide qualitative information on recharge conditions.}
\change[author]{Generally, more excess air is seen in wells located in the center of the plain. This may result from water table fluctuations (Aeschbach et al., 2002) between pre- and post-monsoon which are more pronounced in the central plain than in the East and West (Friesen et al., 2018). In addition, t}{T}he central wells at the foot of the mountains (20, 25, 33) exhibit less $^{4}$He$_\mathrm{rad}$ than wells downstream (19) or in the Western (66) plain\change[author]{, indicating younger water.}{ which is attributed to the higher throughflow from the mountains and thus probably shorter residence times in this area.} In contrast, well 1 in the Eastern plain exhibits $^{4}$He$_\mathrm{rad}$ levels two orders of magnitude higher than the rest of the wells and is therefore plotted on a different scale. \add[author]{\protect The high helium value in this well is confirmed by a previous measurement and possibly due to geological factors. In West-East direction the geologic formations are more elevated in the East with the cretaceous formation outcropping at the surface in the area of well 1 \cite{Platel1987, Platel1992}. Other formations are thus flowed through and there is also a poorer connection to the mountain aquifer as the flow rates are lower compared to the central zone \cite{Friesen2018}.}\\
As the $^{4}$He$_\mathrm{rad}$ accumulation rate \change[author]{is}{and its spatial and temporal variability are} unknown, those results only serve for qualitative analysis and will not be further employed in this paper.

\add[author]{\protect\textbf{Tritium:} The measured $^3$H concentrations are hardly significantly different from zero and vary only in a small range comparable to the measurement uncertainty, therefore do not allow reliable differentiation between the wells (Table DS01). The use of tritium as dating tracer in the study area is further hampered by the absence of any stations with precipitation data from the GNIP network \cite{IAEA/WMO2021} in the entire region. Even the few far away stations available around the Arabian Sea and Persian Gulf (Bahrain, Karachi, Mumbai) have only incomplete data. Thus, it is impossible to construct a reliable tritium input function for the entire relevant time including the bomb peak period and especially the last decades. The very few available data for the past 30 years suggest very low $^3$H contents of at most a few tritium units in precipitation at coastal stations around the Arabian Sea. The low measured $^3$H contents in some wells - if at all significant - can therefore not distinguish between remnants from the bomb peak period and fairly recent recharge. For those reasons, tritium measurements were excluded from analysis.}

\textbf{CFC-11 vs. CFC-12:} In Figure \ref{fig_2} b), the measurements of CFC-12 are plotted against CFC-11. In addition, the possible relationships between the tracer concentrations for all mean residence times were modelled for the Piston-flow Model, the Exponential Model and the Dispersion Model (with Péclet number $Pe = 10$) using the groundwater age modelling program \textit{Lumpy} \cite{Suckow2012}.\remove[author]{For this purpose, input parameters similar to those for the $^4$He analysis were chosen.} From visual interpretation, \change[author]{the wells cannot}{not all wells can} be described by the applied models. While most wells show a high level of agreement with respect to their CFC-11 and CFC-12 data, the measurements for wells 19 and 25 are incompatible.
\add[author]{\protect Potential unconsidered excess air would have a larger impact on CFC-12 than on CFC-11 due the lower solubility and could thus explain a deviation towards higher CFC-12. However, the moderate amounts of excess air derived from the noble gas measurements correspond to only small (few percent) effects for all CFCs. In contrast, the results at hand could arise from either a depletion of CFC-11 or a (small) contamination with CFC-12. While all CFCs are susceptible to contamination of anthropogenic origin \cite{Darling2012, IAEA_CFCs, Morris2006, Oster1996}, microbial degradation of CFC-11 in groundwater occurs only under anaerobic conditions \cite{Khalil1989, Oster1996}. However, all wells except well 19 show substantial concentrations of dissolved oxygen (5 to 6.5\,mg/L, compare Table DS01) and can be considered aerobic. Only for well 19 with its lower oxygen level of 0.85\,mg/L, an effect of anaerobic degradation appears possible. CFC-11 degradation is therefore only a plausible explanation for this single well, whereas indications for contamination are also present for CFC-113 and SF$_6$, especially for wells 19 and 25 (see below). Moreover, for several wells the CFC-12 data are not in agreement with the tracers used for longer timescales, so a modelling approach appears only feasible using CFC-11.}
\remove[author]{The results at hand can arise from either a depletion of CFC-11 or a (small) contamination with CFC-12. While all CFCs are susceptible to contamination of anthropogenic origin (IAEA, 2006; Morris et al., 2006; Oster et al., 1996; Darling et al., 2012), microbial degradation of CFC-11 in groundwater occurs only under anaerobic conditions (Oster et al., 1996; Khalil \& Rasmussen, 1989). Although well 19 shows lower oxygen levels than the other wells (compare Table DS01), the conditions of all wells can be considered aerobic. An additional argument emphasizing the credibility of CFC-11 as a tracer is its higher solubility leading to a lower susceptibility to unconsidered excess air. An underestimation in excess air would hence produce a larger error in CFC-12 than in CFC-11. Moreover, for several wells the CFC-12 data are not in agreement with the tracers used for longer timescales, so a modelling approach is only feasible using CFC-11. Hence, CFC-12 data are disregarded for the multi-tracer modelling approach.}

\textbf{CFC-11 vs. CFC-113:}
In Figure \ref{fig_2} c), CFC-113 is plotted against CFC-11. For all wells except for well 66, the CFC-113 measurement data cannot be produced by any of the transit time distributions. This shows that most of the wells are contaminated with respect to CFC-113 and this tracer cannot be further applied for the model.

\textbf{CFC-11 vs. SF\textsubscript{6}: } In Figure \ref{fig_2} d), the CFC-11 and SF\textsubscript{6} data can be seen. The SF\textsubscript{6} concentrations of wells 19 and 25 are above atmospheric levels for any mean residence times, which can only result from a contamination or subsurface production in the central plain. Likewise, SF\textsubscript{6} data will not be further used in this paper.

\begin{figure*}
    \centering
    \includegraphics[width=\linewidth]{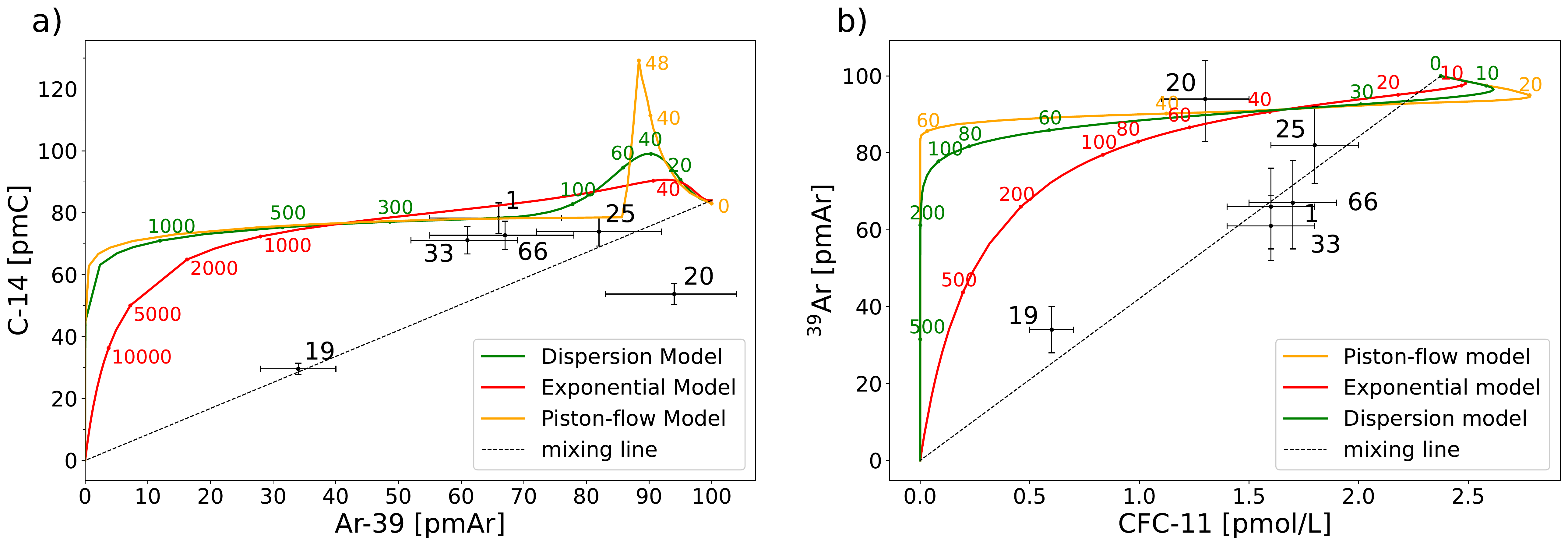}
    \caption{\change[author]{a) Synoptic plot of \textsuperscript{39}Ar and \textsuperscript{14}C data and b) synoptic plot of CFC-11 and \textsuperscript{39}Ar data, both indicating mixing of two water masses.}{Synoptic plots of a) \textsuperscript{39}Ar and \textsuperscript{14}C data as well as b) CFC-11 and \textsuperscript{39}Ar data, both indicating the mixing of two water masses. $^{14}$C was corrected by the Vogel model, resulting in a factor of 0.8 applied to the TTD trajectories.}}
    \label{fig_3}
\end{figure*}

\subsection{Tracers involved in the modelling approach}
After considering the qualitative information of radiogenic \textsuperscript{4}He, CFC-12, CFC-113 and SF\textsubscript{6}, the\add[author]{ dating} tracers remaining for a Bayesian modelling approach are \textsuperscript{14}C, \textsuperscript{39}Ar and CFC-11.\add[author]{\protect  Of these tracers, \textsuperscript{14}C is the most complex to interpret as it is necessary to account for the dissolution of old carbon in groundwater.}
In literature, different correction methods for \textsuperscript{14}C are found\add[author]{, \protect as reviewed by \citeA{Han2016}. A simple model relying entirely on chemical data, especially the alkalinity measurements shown in Table DS01 (Supporting Information), is the Tamers model. For near-neutral pH values as found in our samples (Table DS01), this model yields initial \textsuperscript{14}C activities close to 50\,pmC, which would represent an overcorrection for all wells except for wells 19 and 20. Two widely used models are the the Pearson model, which relies solely on the $\updelta ^{13}$C data of the samples (compare Figure S1 in the Supporting Information), and the Fontes-Garnier model, which combines chemical and isotopic data. Those models require $\updelta ^{13}$C values of the end members soil air (modern component with 100\,pmC) and calcite (old component with 0\,pmC). For the soil air component, the mean (and its error) of four $\updelta ^{13}$C values measured in the infiltration area were used, giving $\updelta ^{13}$C$_\mathrm{soil} = (-17.6 \pm 2.9)$\,\textperthousand\;\cite{Al-Mashaikhi2011}. For the carbonate end member, $\updelta ^{13}\mathrm{C}_\mathrm{calcite} = (2 \pm 1)\,$\textperthousand\;was assumed. However, with these assumptions both models fail due to overcorrection and produce negative apparent ages for all samples except well 19 (Pearson) and wells 19 and 66 (Fontes-Garnier). Overcorrection could be partly avoided by using more positive $\updelta ^{13}$C values for the end members, but we consider that highly unlikely given that the assumed values are experimentally verified and already rather high in comparison to standard assumptions (e.g., -23 \textperthousand\;for $\updelta ^{13}$C$_\mathrm{soil}$ and 0\,\textperthousand\;for $\updelta ^{13}$C$_\mathrm{calcite}$). Other models discussed by \citeA{Han2016} correctly accounting for isotope exchange with soil CO$_2$ under open conditions (Mook's and Han \& Plummer's model) work for more samples, but still leave wells 20 and 25 overcorrected. We therefore assume that our measured $\updelta ^{13}$C values are either not reliable or affected by additional processes that we cannot model}.
\remove[author]{While the Pearson model relies on the 13C data of the samples (compare Figure S1 in the Supporting Information), the Fontes-Garnier and Tamers model are based on alkalinity measurements shown in Table DS01 (Supporting Information). However, those models fail due to overcorrection and produce negative apparent ages.} 
Accordingly, the correction method of choice is the Vogel model with the concentration of modern \textsuperscript{14}C set to $A_0 = 80 \pm 5$\,pmC due to the karstic aquifer environment in the Salalah Plain \cite{Flint1986, Vogel1968}. \add[author]{This simple model with its uniform correction reflects the differences in the measured $^{14}$C activities, indicating substantially higher ages for wells 19 and 20 compared to the other samples. It may, however, lead to somewhat overestimated ages for these two wells by not accounting for their relatively high $\updelta ^{13}$C values (compare Figure S1 in the Supporting Information). }

A tracer-tracer plot of \textsuperscript{14}C and \textsuperscript{39}Ar is shown in Figure \ref{fig_3} a), together with the calculated trajectories of different transit time distributions. \add[author]{While the $^{39}$Ar input curve was assumed constant at \protect 100\,pmAr, the bomb peak from the Northern Atmospheric $^{14}$C input curve was considered.} Only wells 1 and 25 seem to be consistent with the depicted models whereas all other wells clearly cannot be represented by a single water mass. However, the figure suggests the possibility of a mixed water with multiple components of different origin, whereas wells 19 and 20 may still pose difficulties. This observation is corroborated in Figure \ref{fig_3} b) where \textsuperscript{39}Ar and CFC-11 data are shown. For most wells, neither of the one-component age distributions can explain the measurements whereas with a mixing line of two or more water masses, this would be feasible.

\subsection{Nonparametric model results}

\begin{figure}
    \centering
    \includegraphics[width=\linewidth]{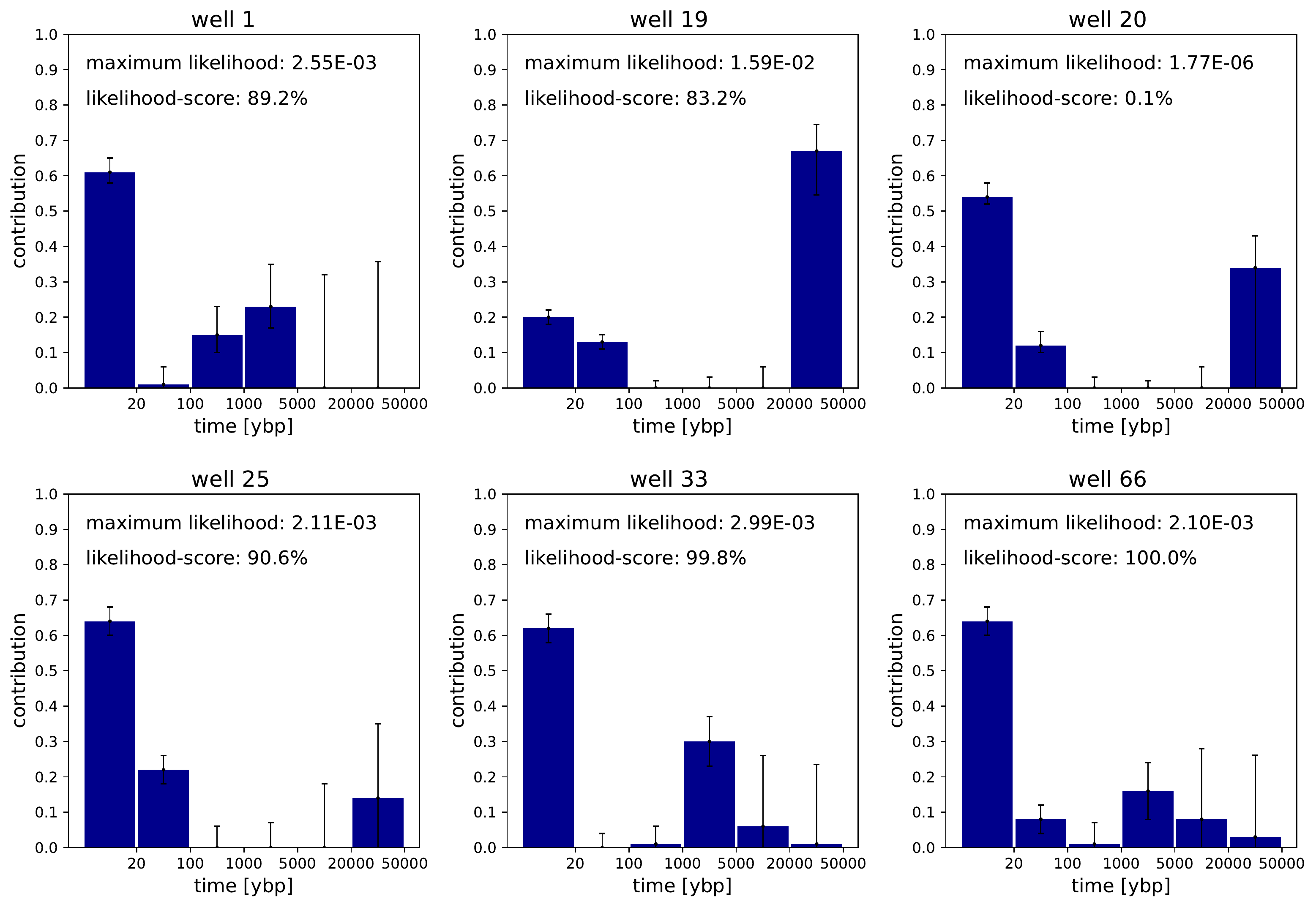}
    \caption{Modelling of the \textsuperscript{39}Ar, \textsuperscript{14}C and CFC-11 tracer data using a nonparametric age distribution with six bins. For all wells, two distinct water masses become apparent.}
    \label{fig_4}
\end{figure}

To allow for a transit time distribution with multiple maxima without imposing an analytic function, tracer concentrations were modeled using a nonparametric approach (as described in Section \ref{meth:math_model}). Six age bins (resulting in five free parameters $\phi_\mathrm{k}$) were selected with boundaries at 20, 100, 1000, 5000, 20000 and 50000 years, implying a constant age contribution within each bin k. Using a Markov-Chain Monte Carlo simulation with 20000 iterations, the best age distribution and the associated maximum likelihood were found. As start parameters, equal groundwater fractions $\phi_\mathrm{k}$ in each bin were chosen. 
As can be seen in Figure \ref{fig_4}, almost all wells show reasonable likelihoods, stating that the measured tracer concentrations may be produced by the displayed age histograms. The reason for the inadequate modelling likelihood of well 20 are the incompatible \textsuperscript{14}C and \textsuperscript{39}Ar measurements (compare Figure \ref{fig_3}). 
The distributions among the age bins show two distinct age clusters: one very young component below 20 years of strong contribution and one old component of varying ages. While for well 66 many bins contribute with small shares, the other wells exhibit a gap between the young and the old cluster. High contributions above 20000 years before present are only found in wells 19 and 20. \add[author]{Those results are visualized as cumulative age distributions in Figure S3 (Supporting Information).}

\subsection{DMmix model results}
Based on the results of the nonparametric model, an analytical model can be designed to find the best representation of the hydrological system. Due to the clustering in very young and very old water, a two-component Dispersion Model (DMmix) with unknown Péclet numbers, mean residence times and unknown mixing ratio of the young contribution is chosen for the further analysis. As those parameters involve large parameter spaces and the analytical model brings along a high computation intensity, it is not possible to model all parameters at once. Instead, a set of Péclet numbers is preselected\add[author]{ ($Pe = 10$)}, reducing the number of free parameters from five to three. \add[author]{\protect A more thorough analysis of the Péclet space is described in the Supporting Information (Text S4 and Figure S4). Hence, only}\remove[author]{ Subsequently,} MRT$_1$,  MRT$_2$ and the mixing ratio \add[author]{$r$ }of water mass 1 \add[author]{(the younger one) }are optimized whereas not only the parameter estimates, but also the likelihood in parameter space is taken into account.

\remove[author]{
Constraining the flow regime for DMmix  

For each water mass, five different Peclet numbers are tested (5, 10, 100, 1000 and 10000). While low Peclet numbers indicate a regime where the Dispersion Model approaches the Exponential Model, high numbers suggesting an advective regime show a tendency towards a Piston-flow. For each of the 25 Peclet combinations, the highest achievable likelihood is calculated. For the Markov-Chain Monte Carlo simulation, a two step approach with a broad sampling in the whole parameter space (10000 iterations) and a more precise, more stringent scanning around the maximum (20000 iterations) are employed. 
FIGURE 5 
The results are depicted in Figure 5. For a more intuitive visualization, the achievable likelihood given the measurement uncertainties (complying with the measured tracer concentrations) is applied as a normalization constant representing 100\% likelihood. As mentioned in the methods section, the resulting quantity of the relative likelihood is denoted as likelihood score, indicated by the color scale. As the respective fit parameters (mean residence times and mixing ratio) are not of interest at this point, they are not displayed in the plot. 
For wells 1, 33 and 66, all Peclet combinations yield likelihoods above 99\,\% with no perceptible trend. Similarly, well 20 shows a continuous low likelihood (around $0.1\,\%$) in Peclet space. For well 19, a propensity for low Pe$_1$ of the young component and for high Pe$_2$ of the old component can be seen, yielding values between 43 and 70\,\%. For well 25, an opposite trend is visible, though not as pronounced (likelihoods between 66 and 70\,\%). Expecting a similar flow regime for all wells, those are conflicting results. In addition, Peclet numbers as high as 10000 indicates a striving towards a Piston-flow Model for one component. For well 19, this effect can be explained by the model's pursuit to dilute the young component with water of the lowest possible $^{14}$C content, to achieve a high level of CFC-11 while satisfying the low $^{14}$C measurements. At this point, an inspection of the tracer reliability is inevitable, which will be attended in the Discussion. 
In literature, the Peclet number in groundwater is usually estimated to be around $Pe = 10$, as was constrained in a scale analysis (Gelhar et al., 1992). As this study aims to design a hydrologically sound model and a Peclet value of 10 does not contradict the modelling outcome displayed in Figure 5, this assumption will be employed for both water masses in the following. 
}

\subsubsection{Maximum likelihood estimates}
\remove[author]{
The remaining parameters to be optimized in the DMmix model are the mean residence times of the two distributions $\mathrm{MRT}_{1}$ and $\mathrm{MRT}_{2}$ as well as the mixing ratio $r$ designating the share of water mass 1 (the younger one). }Using MCMC, the maximum likelihood estimates are computed as described above and are depicted in Table \ref{tab:DMmix_estimates}. \add[author]{The cumulative age distributions are shown in Figure S3 (Supporting Information). }All wells except for well 20 show likelihood scores above 50\,\%, with wells 1, 33 and 66 almost perfectly matching the measured tracer concentrations. On the other hand, the highest maximum likelihood $\mathcal{L}(t \mid \hat{\theta})$ is found for well 19. This effect can be attributed to the smaller measurement uncertainties for lower \change{measurements}{concentrations} (compare Table DS01), resulting in a narrow and high likelihood distribution. This emphasizes the importance of the likelihood score as an additional measure. In general, the mediocre likelihood scores of wells 19 and 25 stem from high $^{39}$Ar measurements suggesting younger water than assumed from $^{14}$C and CFC-11 measurements (compare Table DS01). This is manifested most distinctly for well 20 where the likelihood score is below 1\,\%. \add[author]{A possible explanation for the disagreement between the tracers could be an undercorrection of the $^{14}$C ages for these wells with relatively high $\updelta ^{13}$C values, which would require further geochemical investigations to resolve.}\\
The parameter estimates of MRT$_1$ are divided in two clusters with one being at the lower end of the parameter space below 10\,y (wells 20, 25 and 33) and one between 27\,y and 32\,y (wells 1, 19 and 66). Likewise, MRT$_2$ allows a classification in wells below 10000\,y (1, 25, 33), wells in the middle range (66) and wells bordering the constraints at 40000\,y (19, 20). Unfortunately, the clusters based on the first two parameters do not coincide. Hence, an enormous gap between the young and the old component is present, with none of the wells exhibiting mean residence times between 100 and 1000 years.
The young water mass accounts for the main component with a mixing ratio between 0.63 and 0.8, only well 19 (located downstream from the central wells 20, 25 and 33) exhibits a dominant old component (\change[author]{ratio}{young water ratio of} 0.3). \\
The upper and lower errors of the parameter estimates correspond to the confidence intervals of 68.27\,\%, which would comply with the 1$\upsigma$ interval for a normal distribution. In several cases, only a lower bound for the uncertainty can be given due to parameter constraints. It becomes clear that the uncertainties of MRT$_2$ are significantly larger than those of MRT$_1$ and the ratio. However, the confidence intervals only serve as a conservative estimate. 

\begin{table}
\caption{Parameter estimates and their likelihood for the DMmix model.}
\renewcommand{\arraystretch}{1.5}
\centering
\begin{tabular}{ c  c  c  c  c  c }
	\toprule
	well & MRT$_1$ [ybp] & MRT$_2$ [ybp] & mixing ratio $r$ & max lik & max lik-score [$\%$]\\ 
	\midrule
	1 & $26.9 \substack{+9.7\\->23.9^*} $ & $4414 \substack{+8682\\-3557} $ & $0.678 \substack{+0.115\\-0.091}$ & 535.9 & 100.00 \\
	19 & $32.3 \substack{+10.9\\-11.7}$ & $40000 \substack{+>0\\-13355}$ & $0.297\substack{+0.050\\-0.030}$ & 1885.4 & 52.62 \\
	20 & $3.0 \substack{+9.4\\->0.0}$ & $40000 \substack{+>0\\-25422}$ & $0.674 \substack{+0.094\\-0.068}$ & 0.8 & 0.14 \\
	25 & $3.0 \substack{+17.8\\->0.0}$ & $6346 \substack{+>33654\\->6341}$ & $0.811 \substack{+0.100\\-0.102}$ & 295.0 & 67.73 \\
	33 & $9.6 \substack{+25.1\\->6.6}$ & $4249 \substack{+17907\\-683}$ & $0.626 \substack{+0.119\\-0.043}$ & 561.9 & 100.00 \\
	66 & $29.0 \substack{+7.3\\-12.6}$ & $22497 \substack{+>17503\\-17016}$ & $0.753 \substack{+0.074\\-0.085}$ & 393.1 & 100.00 \\
   	\bottomrule
	\multicolumn{6}{l}{$^{*}$ For error margins exceeding the parameter bounds, a lower threshold is given.}
\end{tabular}
\label{tab:DMmix_estimates}
\end{table}

\subsubsection{Likelihood score in parameter space}

\begin{figure*}
    \centering
    \includegraphics[width=.8\linewidth]{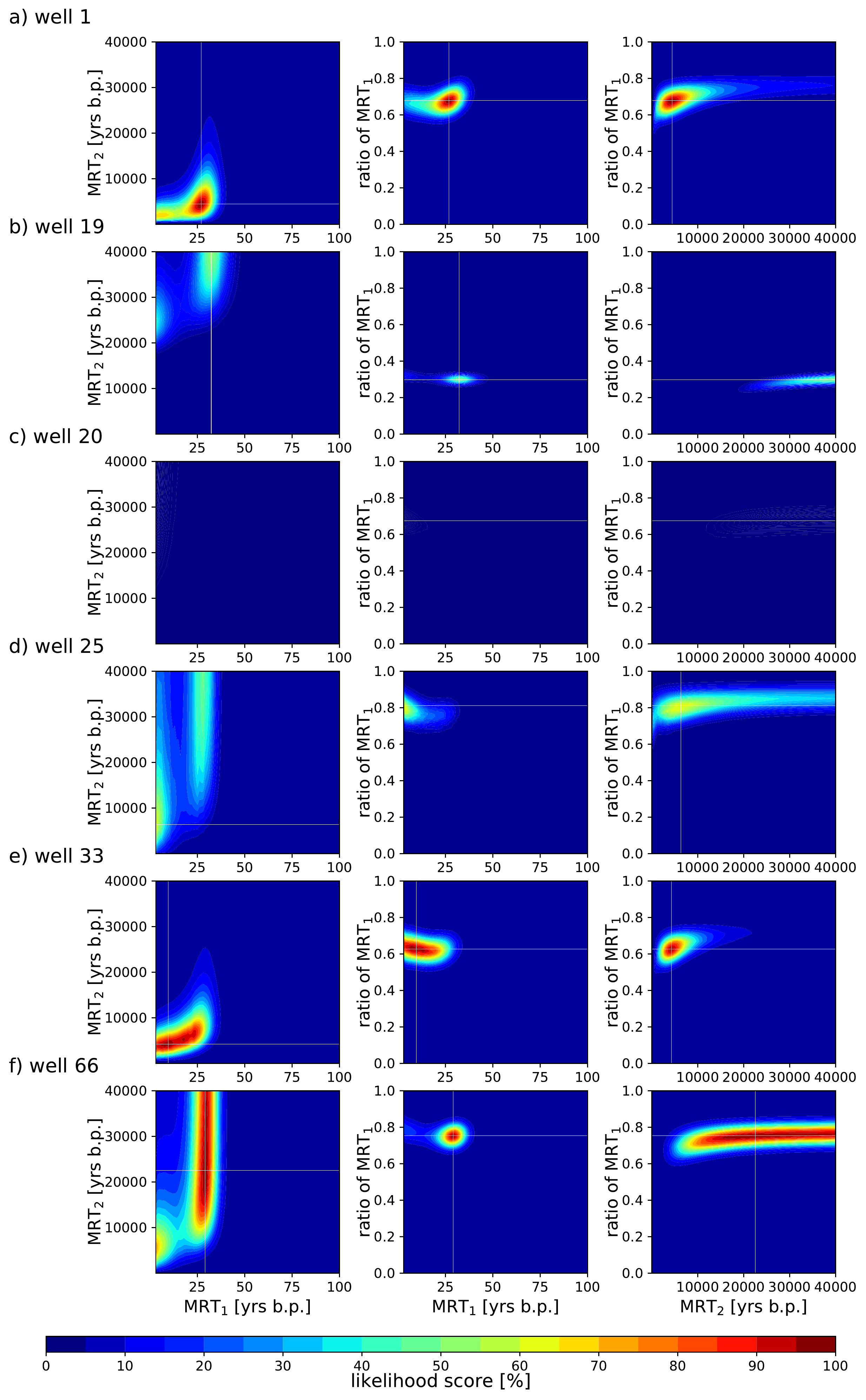}
    \caption{Likelihood score of all wells in parameter space with lines through the maximum. Only wells 1, 33 and 66 achieve $>95\,\%$. The spread along the axes indicates the magnitude of the uncertainties of the estimates.}
    \label{fig_6}
\end{figure*}

While the MCMC algorithm is designed to find the point of maximum likelihood, a thorough uncertainty analysis is required to understand error margins and codependencies of the parameters. Consequently, the entire distribution along the parameter axes is shown in Figure \ref{fig_6}. The values were calculated by screening along parameter surfaces intersecting the point of maximum likelihood, with step sizes $\Delta \mathrm{MRT}_{1} = 1$\,y,  $\Delta \mathrm{MRT}_{2} = 200$\,y and $\Delta r = 0.01$. The points of maximum likelihoods are marked with crosshairs and correspond to the parameter estimates listed in Table \ref{tab:DMmix_estimates}. 

In the left plots where both MRTs are contrasted, two maxima are visible for most wells. Presumably, this stems from the non-monotonic CFC-11 input curve yielding too high concentrations for $5$\,y$ < \mathrm{MRT_1} < 20$\,y which can be verified in the Dispersion Model trajectories for CFC-11 in Figures \ref{fig_2} and \ref{fig_3}. This effect explains the clustering of the estimates for MRT$_1$ mentioned above. \\
The maxima of the likelihood score in parameter space are well-defined regarding MRT\textsubscript{1} and the ratio whereas MRT\textsubscript{2} has very large tolerance towards the upper age end (revealing the slow exponential decay of \textsuperscript{14}C, the only tracer applicable in this regime). This is most evident for well 66 where any value for MRT$_2$ between 10000 and 40000 years yields a high likelihood score. While this appears favorable, the associated error boundaries are respectively large. Worse still for well 25, the blurred likelihood profile results in error margins spanning the entire parameter range of MRT$_2$ (compare Table \ref{tab:DMmix_estimates}). For well 20, the likelihood score does not reach a high enough level to be resolved in the applied color scale. However, the contour lines and the crosshairs indicate the location of the maximum. \\
For most wells, only parts of the parameter space exhibit notable likelihoods. However, the \change{constraints}{boundaries} were not adjusted individually to allow for comparison between the wells. This impacts the model's performance which is discussed in the following section. 

\subsection{Model comparison}

The findings of the nonparametric model given the predefined bin boundaries are generally in agreement with the DMmix parameter estimates. In addition to evaluating the tracer data, quantitative measures are required for an adequate model comparison. Therefore, several indicators regarding the models are \change[author]{opposed}{contrasted} in Table \ref{tab:model_comparison}.

\subsubsection{Maximum likelihood score}
One measure to evaluate the models' performance is the maximum likelihood as well as the score, which on average is slightly higher for the nonparametric model than for DMmix. For wells 20, 33 and 66, the results of the applied models are almost identical. While both models reproduce the CFC-11 values of the three remaining wells almost perfectly, their performance diverges in respect to the $^{14}$C and $^{39}$Ar values. Looking at the $^{14}$C to $^{39}$Ar relationship, it becomes apparent that the nonparametric model struggles in reproducing a high ratio (well 1) and the DMmix model in reproducing a low ratio (wells 19 and 25). Possible reasons for the different behaviour may be a different weighting of the $^{14}$C bomb peak or the discrete steps of the nonparametric model between bins. Another explanation is the ability of the nonparametric model to produce lower $^{14}$C concentrations due to the chosen model constraints and with that, a more efficient dilution of the entire sample with $^{14}$C-dead water. In addition, wells 19 and 25 are the only two wells for which the choice of Péclet number 10 for the DMmix model may hinder the maximum likelihood (compare Figure S4 \add[author]{in the Supporting Information}). Nevertheless, for reasons of simplicity and consistency, the same Péclet numbers were chosen for all wells and components.

\subsubsection{Bayes Factor}
The second quantity to be analyzed is the Bayes factor $\ln{B_\mathrm{k}}$, a measure for the mean likelihood in parameter space. For all wells, the Bayes factor for DMmix is slightly higher than for the nonparametric model, indicating that DMmix performs better (compare Table \ref{tab:model_comparison}). For both models and most wells, the average log-likelihood is between 0.5 and 2. Only well 20 yields a considerably lower Bayes Factor, as presumed from Figure \ref{fig_6}. 
A good Bayes Factor is achieved for models with a clever prior selection (or, in this study, an advised choice of the parameter constraints). Hence, the 5-dimensional parameter space of the nonparametric model harbors more sectors of low likelihood than the 3-dimensional parameter space of DMmix. However, under the perspective of equity, it is essential to mention that the parameter constraints of the mean residence times in DMmix were designed based on results of the nonparametric model. In particular, MRT$_1$ with its upper limit of 100\,y is able to avoid vast stretches of redundant parameter space, making possible a faster computation in MCMC, a higher precision and, manifestly, a better Bayes Factor.
Comparing the different wells, well 19 yields the best results for the DMmix model while for the nonparametric model, it is only ranked fourth after wells 33, 66 and 1. This effect again displays the higher likelihood for lower measurement uncertainties (compare Table DS01) and contrasts the impression from Figure \ref{fig_6} where the likelihood score is illustrated.

\subsubsection{Leave-one-out cross validation}
The third quantity listed in Table \ref{tab:model_comparison} is the expected log likelihood for leave-one-out (LOO) cross validation. With $\ln{\hat{\mathcal{L}}_\mathrm{LOO}}$, overfitting is penalized by making predictions on data not used during the training of the model. 
Analogous to the Bayes Factor, the $\ln{\hat{\mathcal{L}}_\mathrm{LOO}}$ results of both models are in the same order of magnitude, but with an opposite trend; more precisely, the nonparametric 6-bin model performs better than the DMmix model for all wells. This is surprising as the nonparametric model involves a higher number of free parameters, a key quantity applied to estimate overfitting in other model performance approaches. This reveals that for this study, the number of free parameters is not the pivotal measure to rate the models' predictive performance. On the contrary, the DMmix model seems to be more capable to tailor a water mixture producing specific tracer concentrations while the nonparametric model is bound to its uniform age contributions within each bin, hindering its flexibility with regard to far-fetched model outcomes. As a result, the likelihood in parameter space produced with and without one of the tracers are more in accordance with each other for the nonparametric model.

Among the wells, the DMmix model ranks well 66 first while for the nonparametric model, well 25 performs best. Besides the exact order of the well-performing wells, a similar trend is visible for both models with the $\ln{\hat{\mathcal{L}}_\mathrm{LOO}}$ of well 19 being significantly lower and well 20 taking the last place. This result is not unexpected as for both wells, the $^{14}$C and $^{39}$Ar results are not in accordance. For incompatible tracer measurements, a procedure based on leaving out one tracer at a time and evaluating the result with respect to this tracer will evidently distort the image more drastically and produce poor likelihoods. 
Additionally, LOO results can also be visualized to see the effect of the individual tracers on the parameter estimates (compare Figure S5 in the Supporting Information).

\begin{table}
\caption{Comparison of the maximum likelihood, the maximum likelihood score, the Bayes factor and the leave-one-out-likelihood for the DMmix and the nonparametric model.}
\begin{tabular}{c c c c c c c c c}
	\toprule
	\multicolumn{1}{c}{Well} & 
	\multicolumn{2}{c}{max. lik $\mathcal{L}(\hat{\theta})$} & 
	\multicolumn{2}{c}{max. lik score [$\%$]} &
	\multicolumn{2}{c}{Bayes factor $\ln{B_\mathrm{k}}$} & 
	\multicolumn{2}{c}{$\ln{\hat{\mathcal{L}}_\mathrm{LOO}}$} \\
	\cmidrule(lr){2-3}
	\cmidrule(lr){4-5}
	\cmidrule(lr){6-7}
	\cmidrule(lr){8-9}
	& DMmix & nonpar & DMmix & nonpar & DMmix & nonpar & DMmix & nonpar \\
	\midrule
	1 & 536 & 478 & 100.0 & 89.2 & 1.641 & 1.236 & 2.02 & 2.38 \\
	19 & 1885 & 2983 & 52.6 & 83.2 & 2.027 & 0.758 & 0.34 & 0.87 \\
	20 & 0.79 & 0.33 & 0.1 & 0.1 & -6.638 & -8.465 & -15.62 & -12.92  \\
	25 & 295 & 395 & 67.7 & 90.6 & 1.382 & 0.594 & 2.49 & 3.88 \\
	33 & 562 & 561 & 100.0 & 99.8 & 2.001 & 1.999 & 2.88 & 3.57 \\
	66 & 393 & 393 & 100.0 & 100.0 & 1.939 & 1.596 & 3.00 & 3.73 \\
	\bottomrule
\end{tabular}
\label{tab:model_comparison}
\end{table}

\section{Discussion}

\subsection{Intercomparison of the findings}
While several tracers measured in the campaign were not applied for Bayesian modelling, their qualitative evidence can be compared with the model outcomes. As all measurements of CFC-113 and several SF$_6$ concentrations indicate contamination in the respective wells, no relevant findings arise from those tracers worth comparing with modelling results. For CFC-12, the parameter estimates attained in the modelling approach could be applied to calculate the CFC-12 concentrations and tracer likelihoods which are displayed in Table S2 (Supporting Information). As expected based on the CFC-12 vs. CFC-11 plot in Figure \ref{fig_2}, for most wells CFC-12 results generally convey a similar picture with slightly different likelihoods. Wells 19 and 25 form an exception as they yield poor CFC-12 likelihoods due to the large difference between CFC-11 and CFC-12 concentrations.

From $^4$He$_\mathrm{rad}$ results, well 1 would be inferred the oldest well by far (compare Figure \ref{fig_2}). Instead, well 1 exhibits the youngest mean residence times (or higher ratios in young age bins for the nonparametric model). The high radiogenic helium at well 1 can possibly be explained by the geology. The coastal plain is strongly characterized by uplift and erosion, visible, for example, on the different layers that wedge out on the surface. In the area of well 1, Cretaceous layers are exposed, while Paleocene layers are pending towards the West \cite{Khalifa1988}. \change[author]{Hence,}{The low measured $^{3}$He/$^{4}$He ratio of 0.08\,Ra in well 1 indicates a purely crustal He origin, which could be due to locally enhanced U and Th concentrations in the rocks or the presence of a fault channeling the crustal He flux at this location. In either case, it is likely a local phenomenon and} the enigma of repeated excessive radiogenic helium measurements in the Eastern Salalah Plain cannot be solved in this study. However, the impression that wells 25 and 33 are among the youngest is supported by the modelling results.

\subsection{Critical assessment of the applied methods}
 
In the context of the model, up to five parameters estimates are calculated with only a set of three tracers. To assure a sufficiently determined system, it is essential to evaluate results from repeated execution with respect to their reproducibility. While the optimization process with six bins (five free parameters) in the nonparametric model proves to be reliable, increasing the number of bins to seven bins (six free parameters) leads to poor reproducibility of the estimates. Hence, the choice of a six-bin model arises from the highest age resolution under simultaneous consideration of a reliable estimation process. For the DMmix model, however, a possible underdetermination was prevented by reducing the parameters to only MRT$_1$, MRT$_2$ and the ratio by separately evaluating the two Péclet numbers. \\
To reduce the risk of contamination, predominantly wells located along the front of the Dhofar mountains upstream of agricultural and residential areas were chosen. Only well 19 poses the danger of infiltrated irrigation water as it is situated further downstream amid agricultural area. However, there is no indication of anthropogenic contamination in well 19 and its location further down the flow path is reflected in a higher MRT$_2$ and a lower ratio of the young component, compared to the other wells. For well 20, the very high $^{39}$Ar results contradicting the CFC-11 and $^{14}$C measurements may have been caused by a contamination of the sample. Subsurface production of $^{39}$Ar \cite{Florkowski1992} can be excluded as the Salalah Plain does not exhibit a high granite content \cite{Khalifa1988}. Noteworthy, despite incomplete data on well properties, well 20 seems to exhibit an excessively long screen and the pump was placed near the surface. Hence, despite thorough flushing before sampling, the inconsistent tracer data may be ascribed to stagnant water in the samples. \\
Assessing the credibility of the tracers involved in the Bayesian modelling approach leads to wide differences between $^{14}$C, $^{39}$Ar and CFC-11. For $^{14}$C, the tracer stipulating the old range, all correction methods involving the groundwater chemistry or $\updelta ^{13}$C failed, leaving only the Vogel model as an empirical estimate. Combined with its high reactivity and subjection to fractionation, this does not inspire a high level of confidence for this tracer which was taken into account by assigning an uncertainty of 5\,\% to the Vogel correction. \add[author]{On top of this uncertainty, it is possible that an appropriate $\updelta ^{13}$C-based correction would substantially reduce the $^{14}$C ages for wells 19 and 20, thereby reducing discrepancies to the other tracers as well as the contributions of the oldest age bins to the TTDs of these wells. }$^{39}$Ar, on the other hand, is not involved in geochemical processes and no fractionation processes need to be considered, making it the most robust tracer in the study. CFC concentrations may be altered by contamination arising from industrial activities, leaking landfills or sewage lines and effects from irrigation or pesticides \cite{Darling2012, IAEA_CFCs, Oster1996}. In addition, the sampling tubing was stored in contact with atmospheric air, potentially allowing CFCs to diffuse into the material.  These effects affect the individual CFCs differently and with CFC-11, the most trustworthy of the three tracers given the measurement results was chosen. \\
In contrast to their differing reliability, all three tracers contribute to the total likelihood in equal shares. The individual tracer likelihoods are substantially governed by the measurement errors of the tracers, allowing considerably higher deviance between modelled and measured activity of $^{39}$Ar than for the other tracers. This is most evident for well 20 where modelled and measured $^{14}$C activities are similar while for $^{39}$Ar they differ widely, despite reliability considerations. This can be taken into account by quantifying a level of reliability for the individual tracers resulting in weight factors for the tracer likelihoods. In lack of a quantification of the tracer reliability, this step was renounced in this study. \\
In addition to the maximum likelihood achievable within a model, the models' performance can be rated. While the DMmix model yields better results for the Bayes Factor, the nonparametric model proves a better predictive adequacy in leave-one-out cross validation. From a methods perspective, calculation of the Bayes Factor is highly cost intensive in terms of computation time. Worse still is $\ln{\hat{\mathcal{L}}_\mathrm{LOO}}$ as the integral of the posterior and likelihood is evaluated over the whole parameter space. For application in further studies this method is not recommended by the authors if computational power is a limited resource. Furthermore, a leave-one-out procedure may be more suitable for a system with more than three tracers. Most notably, leaving out $^{14}$C leaves a major part of the parameter space unresolved (compare Figure S5 in the Supporting Information), which obviously yields results eminently different to those modelled with all three tracers.

\subsection{Hydrological interpretation}

Taking the parameter estimates as the 'true' representation of the hydrological setting, the results of the sampled wells can be employed to assess the groundwater system in the Salalah Plain. From preliminary studies, groundwater ages in the range of decades up to centuries were expected \cite{Clark_1987}, with slightly lower ages in the central plain due to a higher flow rate \cite{GeoResourcesConsultancy2004}. In this study, however, all wells (excluding well 20 due to poor goodness of fit) yield a conclusive picture of the composition with regard to two distinct water masses of very young ($<30$\,y) and very old origin ($>1000$\,y). The ratio and the exact ages, however, differ widely among the wells, potentially due to the difference of the locations in East-West direction and along flow lines (compare Figure \ref{fig:sampled_wells}). The high ratio of young water could be an indication of a short distance between the infiltration site and the well or fast flow paths, both in combination with only shallow infiltration. During the monsoon and especially during extreme events with large-scale flooding, rain or water that has accumulated in ponds and depressions may infiltrate in the immediate vicinity of the well. Before 2018, Cyclone Keila (2011) was the last storm bringing large amounts of rain to the Arabian South coast and flooding large parts of the plain for several days \cite{Strauch2014}. Besides seasonal or extreme events, there is a continuous inflow from the mountains into the coastal aquifer driven by more than 500\,m difference in groundwater head. In this baseflow, mixing of water with short and long flow paths takes place, especially at the outlet - the transition from mountain aquifer to coastal aquifer where the thickness of the water bearing horizon decreases.

\section{Conclusion}
In this study the groundwater system of the Salalah Plain in Southern Oman was investigated by modelling transit time distributions based on a Bayesian multi-tracer approach. The results show that it is possible to model the measured concentrations of $^{14}$C, $^{39}$Ar and CFC-11 with TTDs of different shape using Bayesian statistics. The applied models, a two-component Dispersion Model (DMmix) and a shape-free nonparametric model with six bins, were compared with respect to their maximum likelihood, the Bayes Factor and the leave-one-out expected log likelihood. On average, the nonparametric model performed slightly better in terms of maximum likelihood and leave-one-out cross validation while the Bayes Factor was minimally lower than in the DMmix model. All wells exhibited age distributions of two clusters with a large gap in between, indicating water below 30\,y and beyond 1000\,y. While the occurrence of two peaks in the age distribution of groundwater has been found in other studies as well \cite{Broers2021, McCallum2017}, the extreme age difference observed here seems unusual and may be related to the rather special hydrogeological and climatic conditions of the Salalah Plain. 

Those findings play a crucial role in the estimation of the renewal rate of the groundwater in the Salalah Plain which is currently abstracted in high amounts and constitutes an indispensable resource for agricultural production. To close the gaps of this study, more sampling campaigns in the study area including soil characteristics for a more evidence-based $^{14}$C correction as well as a deeper understanding of the well structure related to the aquifer are required. In addition, the MCMC methods may be replaced by a faster strategy like evolutionary algorithms \cite{Holland1992, Turing1950} to allow for a wider applicability of the Bayesian modelling software in other environmental studies.
\newpage

\nocite{Al-Mashaikhi2011, Gelhar1992}

\acknowledgments
This work was supported by the Deutsche Forschungsgemeinschaft in the joint projects AE 93/14-1 and OB 164/11-1, as well as the project 'Submarine Groundwater Discharge: Adaption of an Autonomous Aquatic Vehicle for Robotic Measurements, Sampling and Monitoring', funded by The Research Council of Oman (TRC Research Contract No. TRC/RCP/15/001). We kindly acknowledge steady support by the Ministry of Agriculture, Fisheries and Water Resources of the Sultanate of Oman. We thank Fynn Bachmann, Dominik Lorenz and Michael Aichmüller for useful discussions.

%\bibliography{../Literatur/Oman_Paper}
\bibliography{Bibliography_RXXdle2022}
\end{document}

\graphicspath{{../Plots/}}
\addeditor{author}

\makeatletter 

\renewcommand{\thefigure}{S\@arabic\c@figure}
\makeatother

\makeatletter 
\renewcommand{\thetable}{S\@arabic\c@table}
\makeatother

%
%  Uncomment the following command to allow illustrations to print
%   when using Draft:
\setkeys{Gin}{draft=false}

\authorrunninghead{RÄDLE ET AL.}
\journalname{Water Resources Research}
% Shorter version of title entered in capital letters:
%\titlerunninghead{SHORT TITLE}

%\authoraddr{Corresponding author: Viola Rädle, 
%Kurt-Eisner-Straße 28, 04275 Leipzig, Germany (viola.raedle@web.de)}

%\begin{document}

\includegraphics{agu_pubart-white_reduced.png}

\title{Supporting Information for 'Multi-Tracer Groundwater Dating in Southern Oman using Bayesian Modelling'}

%DOI: 10.1002/%insert paper number here%

\authors{Viola Rädle\affil{1,5}, Arne Kersting\affil{1}, Maximilian Schmidt\affil{1,2}, Lisa Ringena\affil{2}, Julian Robertz\affil{2}, Werner Aeschbach\affil{1}, Markus Oberthaler\affil{2}, Thomas Müller\affil{3,4}}

\affiliation{1}{Heidelberg University, Institute of Environmental Physics}
\affiliation{2}{Heidelberg University, Kirchhoff-Institute for Physics}
\affiliation{3}{Helmholtz-Centre for Environmental Research, Leipzig}
\affiliation{4}{Helmholtz Centre for Ocean Research, GEOMAR, Kiel}
\affiliation{5}{Leipzig University of Applied Sciences (HTWK)}

%\begin{article}

\noindent\textbf{Contents of this file}
\begin{enumerate}
\item Measurements and well data (Text and Dataset S1)
\item Carbon isotopes (Figure S1)
\item Modelled tracer concentrations and likelihoods (Text\change[author]{ and Table}{, Figure and Table} S2)
\item Testing and performance of the MCMC algorithm and synthetic data sets (Text and Table S3)
\item \add[author]{Cumulative age distributions of both models (Figure S3)}
\item \add[author]{\protect DMmix model results in Péclet space (Text and Figure S4)}
\item Leave-one-out likelihood of an example well (Text and Figure S5)
\end{enumerate}

\noindent\textbf{Additional Supporting Information (Files uploaded separately)}
\begin{enumerate}
\item Table S1 'SI Measurement Data' (Excel-File)
\end{enumerate}

\noindent\textbf{Introduction}
In the supporting information, all well data and all measurements of the groundwater study are listed in an external excel file. The carbon isotopes $\updelta ^{13}$C and $^{14}$C are visualized, supporting the inapplicability of the Pearson correction model. The exact modelled tracer concentrations and likelihoods are given, including the tracer Chlorofluorocarbon 12 (CFC-12) that was not involved in the modelling process. \add[author]{The modelled concentrations are compared to the measurements in a separate figure.} In addition, the programming workflow is described, as the Markov-Chain Monte Carlo algorithm was continuously tested on synthetic datasets. \add[author]{\protect The model outcomes, i.e. the cumulative age distributions for the parameter estimates, are shown for both models. Furthermore, a thorough analysis of the DMmix model results in Péclet space is presented.} At last, a Leave-One-Out procedure is visualized for an example well (well 33) by plotting the LOO-likelihood in parameter space.

\clearpage

\noindent\textbf{Text S1 (Measurement data of the sampled wells)}
The measurement data including field parameters and well specifications are displayed in the Table 'SI Measurement Data'. For the $^{39}$Ar data, the maxima of the histograms are given while the upper and lower error margins of the 1$\upsigma$-level were calculated. 
Most $^3$H measurements are below the detection limit. Concentrations higher than atmospheric are indicated in italic (applies for SF$_6$ and CFC-113).

\noindent\textbf{Text S2 (Modelled tracer concentrations and likelihoods)}
In the process of finding the parameter estimates, the concentrations of the tracers $^{14}$C, $^{39}$Ar and CFC-11 are calculated by the respective mathematical models and their log-likelihoods are obtained by comparison to the measured tracer concentrations. The values are listed in Table \ref{tab:tracer_specs}, including CFC-12 which was not taken into account for the optimization procedure. CFC-12 data can be thought of as a Leave-One-Out test dataset, implying the prediction capability of a model (precluding the eventuality of a contamination). \add[author]{In addition, the measured and the modelled concentrations are shown in Figure} \ref{fig:sup_concentrations}.

In analogy to the total likelihood described in the Results section, the tracer concentrations for wells 1, 33 and 66 match the measurements almost perfectly. For well 1 in particular, the nonparametric model is inferior to the DMmix model due to its incapacity to model a high enough $^{14}$C concentration. On the other hand, it becomes apparent for wells 19 and 25 that the DMmix model has difficulties generating a high enough $^{39}$Ar concentration while yielding the required $^{14}$C and CFC-11 levels. Finally, well 20 exhibits unusually high $^{39}$Ar concentrations at the same time as moderate or low $^{14}$C and CFC-11, impossible to model by any of the models, accordingly leading to a poor log-likelihood for $^{39}$Ar. 
The CFC-12 concentrations for wells 1 and 66 produce good results for both models. For well 20, only the nonparametric model yields a reasonable log-likelihood while for well 33, DMmix performs better. For wells 19 and 25, none of the models provides CFC-12 values in the magnitude of the measured concentrations.

\noindent\textbf{Text S3 (Testing of the Markov-Chain Monte Carlo algorithm and synthetic data sets)}
The Markov-Chain Monte Carlo (MCMC) algorithm has to be tested to ensure its capability with respect to time efficiency and accuracy. For this purpose, synthetic data are produced by specifying an example parameter set $\theta_\mathrm{synth}$, determining the corresponding tracer concentrations $c_\mathrm{synth}$ by forward modelling. Those data are fed into the MCMC algorithm, making possible a validation of the parameter estimates $\hat{\theta}$ with the original parameter set $\theta_\mathrm{synth}$. This way, a both accurate and cost effective mode can be found for all models. The results for DMmix and the nonparametric model are shown in Table \ref{tab:MCMCtesting}. It is apparent that in the DMmix model (with presumed Péclet numbers and three free parameters), computation requires less iterations and time but the precision is not impeccable. The lower part of the table is subdivided into the three parameters of DMmix whereas for the nonparametric model, the same conditions apply to all $\theta_\mathrm{k}$. The proposal density in the first step is wider than in the second step for both models, allowing a larger step size in the initial broad scan. The start parameter range for DMmix as well as the constraints of the mean residence times $\tau_1$ and $\tau_2$ were specified based on preliminary results and model performance.

 \noindent\textbf{Text S4 (Constraining the flow regime for DMmix)}

\add[author]{\protect The DMmix model, a two-component Dispersion Model, involves five free parameters (Péclet numbers $Pe_1$ and $Pe_2$, mean residence times $\tau_1$ and $\tau_2$ and the mixing ratio of the young contribution $r$). As it is not possible to model all parameters at once, presumed Péclet numbers of $Pe=10$ are used in the paper. However, valuable information about the flow regime may be drawn from trends in Péclet space. Hence, a more thorough analysis is conducted in this section. } \\
For each water mass of the DMmix model, five different Péclet numbers are tested (5, 10, 100, 1000 and 10000). While low Péclet numbers indicate a regime where the Dispersion Model approaches the Exponential Model, high numbers suggesting an advective regime show a tendency towards a Piston-flow. For each of the 25 Péclet combinations, the highest achievable likelihood is calculated. For the Markov-Chain Monte Carlo simulation, a two-step approach with a broad sampling in the whole parameter space (10000 iterations) and a more precise, more stringent scanning around the maximum (20000 iterations) are employed. \\
The results are depicted in Figure \ref{sup:fig_peclet}. For a more intuitive visualization, the achievable likelihood given the measurement uncertainties (complying with the measured tracer concentrations) is applied as a normalization constant representing 100\% likelihood. As mentioned in the methods section, the resulting quantity of the relative likelihood is denoted as likelihood score, indicated by the color scale. As the respective fit parameters (mean residence times and mixing ratio) are not of interest at this point, they are not displayed in the plot. \\
For wells 1, 33 and 66, all Péclet combinations yield likelihoods above 99\,\% with no perceptible trend. Similarly, well 20 shows a continuous low likelihood (around $0.1\,\%$) in Péclet space. For well 19, a propensity for low Pe$_1$ of the young component and for high Pe$_2$ of the old component can be seen, yielding values between 43 and 70\,\%. For well 25, an opposite trend is visible, though not as pronounced (likelihoods between 66 and 70\,\%). \change[author]{Those are conflicting results given that wells 19 and 25 are located along the same streamlines and a similar flow regime can be expected.}{Expecting a similar flow regime for all wells, those are conflicting results.} In addition, Péclet numbers as high as 10000 indicates a striving towards a Piston-flow Model for one component. For well 19, this effect can be explained by the model's pursuit to dilute the young component with water of the lowest possible $^{14}$C content, to achieve a high level of CFC-11 while satisfying the low $^{14}$C measurements. At this point, an inspection of the tracer reliability is inevitable, which is attended in the Discussion. \\
In literature, the Péclet number in groundwater is usually estimated to be around $Pe = 10$, as was constrained in a scale analysis \cite{Gelhar1992}. As this study aims to design a hydrologically sound model and a Péclet value of 10 does not contradict the modelling outcome displayed in Figure \ref{sup:fig_peclet}, the assumption $P = 10$ is employed for both water masses of the DMmix model\add[author]{ in the paper}.

\noindent\textbf{Text S5 (Leave-one-out Likelihood for well 33)}
In Figure \ref{fig:sup_lik_loo}, the likelihood score in parameter space is shown for an example well (well 33). In Figure \ref{fig:sup_lik_loo} b), c) and d), not all three tracers were considered but only two of them, analogously to Leave-One-Out (LOO) cross validation. Leaving out $^{14}$C, $^{39}$Ar and CFC-11 shows the impact of the individual tracers on the results. 
Without $^{14}$C tracer data, a blur is present in MRT\textsubscript{2} over the entire range, showing that $^{14}$C is responsible for constraining this parameter. MRT$_2$ is higher than the original one. However, due to the lack of variation of the likelihood along the MRT$_2$ axis, no reasonable conclusions can be drawn from this finding. 
Figure \ref{fig:sup_lik_loo} c) reveals that leaving out $^{39}$Ar doesn't affect a specific axis in the likelihood distribution. The point of Maximum Likelihood is found for a MRT$_1$ slightly smaller than the one in Figure \ref{fig:sup_lik_loo} a). In addition, the tail in MRT$_2$ is less defined without the $^{39}$Ar information. 
Leaving out CFC-11 causes a wider acceptance range with respect to MRT\textsubscript{1} towards older values, as depicted in Figure \ref{fig:sup_lik_loo} d). Hence, the young groundwater component is predominantly confined by CFC-11. In the middle plot, two maxima can be seen with a interjacent gap. This shows a clear avoidance of the \textsuperscript{14}C bomb peak prohibiting MRTs around 50 y.b.p whereas older MRTs are allowed. In the right plot the tail in MRT$_2$ is again less restricted than in the Figure \ref{fig:sup_lik_loo} a).

\bibliography{../Literatur/Oman_Paper}

% if you get an error about newblock being undefined, uncomment this line:
%\newcommand{\newblock}{}

%\end{article}

\clearpage

\noindent\textbf{Figure S1 (Carbon isotopes)}
\begin{figure}[H]
\setcounter{figure}{0}
    \centering
    \includegraphics[width=0.7\linewidth]{Figure_S1/Figure_C13C14.pdf}
    \caption{\textsuperscript{14}C vs. $\updelta ^{13}$C data of the wells and end members (modern component: soil air, old component: calcite) with soil $\updelta ^{13}$C values taken from \citeA{Al-Mashaikhi2011}. Only wells 19 and 66 are within the error margins of the mixing line\add[author]{ or below}, showing that the Pearson model cannot be applied to correct for calcite dissolution.}
    \label{fig:sup_c13c14}
\end{figure}

\newpage
\noindent\textbf{Table S2 (Modelled tracer concentrations and likelihoods)}

\begin{table}[H]
\setcounter{table}{1}
\caption{Modelled tracer concentrations and likelihoods.}
\resizebox{\textwidth}{!}{%
\begin{tabular}{c c c c c c c c c c}
	\toprule
	\multicolumn{2}{c}{ } & 
	\multicolumn{4}{c}{Tracer concentrations} & 
	\multicolumn{4}{c}{Tracer log likelihoods} \\
	%\cmidrule(lr){1-2}
	\cmidrule(lr){3-6}
	\cmidrule(lr){7-10}
	Well & Type/& $^{14}$C & $^{39}$Ar & CFC-11 & CFC-12
	 & $^{14}$C & $^{39}$Ar & CFC-11 & CFC-12  \\
	&Model&[pmC]&[pmAr]&[pmol/kg]&[pmol/kg]&&&&\\
	\midrule
	1 &measured& 78.3 & 63.3 & 1.6 & 1.0 &&&& \\
	  & DMmix & 78.3 & 63.3 & 1.6 & 0.9 & 2.50 & 2.17 & 1.61 & 1.62 \\
	  & nonpar & 76.5 & 65.1 & 1.6 & 0.8 & 2.44 & 2.14 & 1.60 & 0.61 \\ 
	\midrule
	19 &measured& 29.6 & 32.7 & 0.6 & 1.1 &&&& \\
	   & DMmix & 29.9 & 27.3 & 0.6 & 0.3 & 3.46 & 1.79 & 2.28 & -26.45 \\
	   & nonpar & 30.1 & 30.7 & 0.6 & 0.3 & 3.44 & 2.27 & 2.29 & -27.86 \\
	\midrule
	20 &measured& 53.8 & 93.9 & 1.3 & 0.9 &&&& \\
	   & DMmix & 56.0 & 66.3 & 1.6 & 0.6 & 2.65 & -3.23 & 0.34 & -2.24 \\
	   & nonpar & 57.2 & 63.0 & 1.5 & 0.8 & 2.36 & -4.68 & 1.22 & 1.37 \\
	\midrule
	25 &measured& 73.9 & 84.0 & 1.8 & 1.4 &&&& \\
	   & DMmix & 74.0 & 79.8 & 1.9 & 0.7 & 2.56 & 1.79 & 1.34 & -20.56 \\
	   & nonpar & 74.4 & 81.4 & 1.8 & 1.0 & 2.55& 1.82 & 1.61 & -7.63 \\	
	\midrule
	33 &measured& 71.1 & 60.9 & 1.6 & 1.1 &&&& \\
	   & DMmix & 71.1 & 61.2 & 1.6 & 1.0 & 2.60 & 2.12 & 1.61 & 1.47 \\
	   & nonpar & 71.1 & 61.0 & 1.6 & 0.8 & 2.60 & 2.12 & 1.61 & -1.36 \\
	\midrule
	66 &measured& 72.7 & 69.6 & 1.7 & 0.9 &&&& \\
	   & DMmix & 72.7 & 69.9 & 1.7 & 1.0 & 2.58 & 1.79 & 1.61 & 1.93 \\
	   & nonpar & 72.7 & 69.8 & 1.7 & 0.9 & 2.58 & 1.79 & 1.61 & 2.08 \\ 
	\bottomrule
\end{tabular}}
\label{tab:tracer_specs}
\end{table}

\newpage
\noindent\textbf{Figure S2 (Modelled tracer concentrations)}
\begin{figure}[H]
\setcounter{figure}{1}
    \centering
    \includegraphics[width=\linewidth]{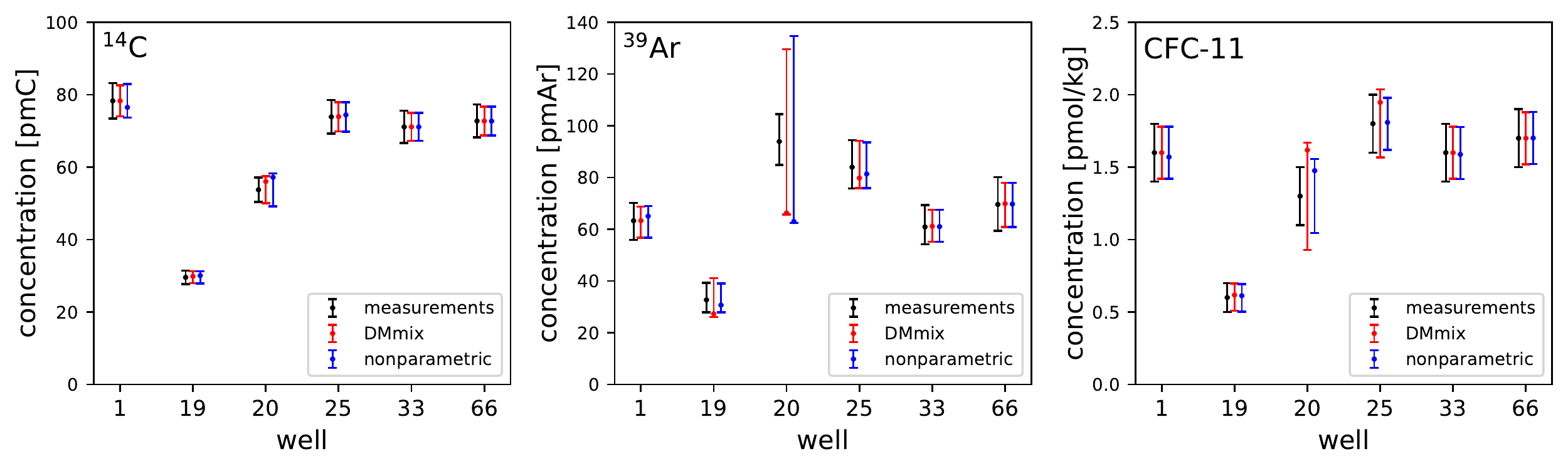}
    \caption{Concentrations of $^{14}$C, $^{39}$Ar and CFC-11 for both models and the measurement data. For wells 1, 33 and 66, the modelled concentrations are almost identical to the measurements. For wells 19 and 25, the DMmix model struggles to reach a high enough $^{39}$Ar value while for well 25, its modelled CFC-11 value is too high. For well 20, both model outcomes yield a much lower $^{39}$Ar value than the measurement and show large upper uncertainties.}
    \label{fig:sup_concentrations}
\end{figure}

\newpage

\noindent\textbf{Table S3 (Testing Markov-Chain Monte Carlo simulation with synthetic datasets)}
\begin{table}[H]
\setcounter{table}{2}
\caption{Properties of the Markov-Chain Monte Carlo simulation for both models.}
\centering
\begin{tabular}{c c c c c}
	\toprule
	Model & nonparametric &  & DMmix & \\ 
	& (6 bins) &&& \\
	\midrule
   	Number of free parameters  & $5$  &  & $3$ & \\
   	Iterations (1st round) & $50000$ &  & $10000$ &  \\
   	Iterations (2nd round) & $10000$ &  & $20000$ &  \\
   	Acceptance (1st round) & $\approx 95\,\%$ &  & $\approx 60\,\%$ &  \\
   	Acceptance (2nd round) & $\approx 4\,\%$ &  & $\approx 10\,\%$ &  \\
   	Precision & $ 100.00\,\%$ &  & $ 99.99\,\%$ &  \\
   	Computation time & $\approx 3\,$min & & $\approx 2\,$min & \\
   	\midrule
   	Parameters & $\theta_\mathrm{k}$ & $\tau_1$ [ybp] & $\tau_2$ [ybp] & $r$ \\
   	\midrule
   	$\sigma_\mathrm{proposal}$ (1st round) & $0.03$ & $5$ & $500$ & $0.1$ \\
   	$\sigma_\mathrm{proposal}$ (2nd round) & $0.02$ & $0.1$ & $10$ & $0.01$ \\
   	constraints & $\theta_\mathrm{k}$ in $\{0, 1\}$, & $\{ 3, 100 \}$& $\{ 3, 40000 \}$ & $\{ 0, 1 \}$ \\
   	&$\sum_\mathrm{k=1}^{5} \theta_\mathrm{k} = 1 - \theta_\mathrm{6}$ & & & \\
   	initial values & $\theta_\mathrm{k} = \frac{1}{n_\mathrm{bins}} = \frac{1}{6}$ & $\{ 3, 30 \}$ & $\{ 1000, 20000 \}$ & $\approx 0.5$ \\
   	\bottomrule
	%\multicolumn{2}{l}{$^{a}$Footnote text here.}
\end{tabular}
\label{tab:MCMCtesting}
\end{table}

\newpage
\noindent\textbf{Figure S3 (Cumulative age distributions of both models)}
\begin{figure}[H]
\setcounter{figure}{2}
    \centering
    \includegraphics[width=.95\linewidth]{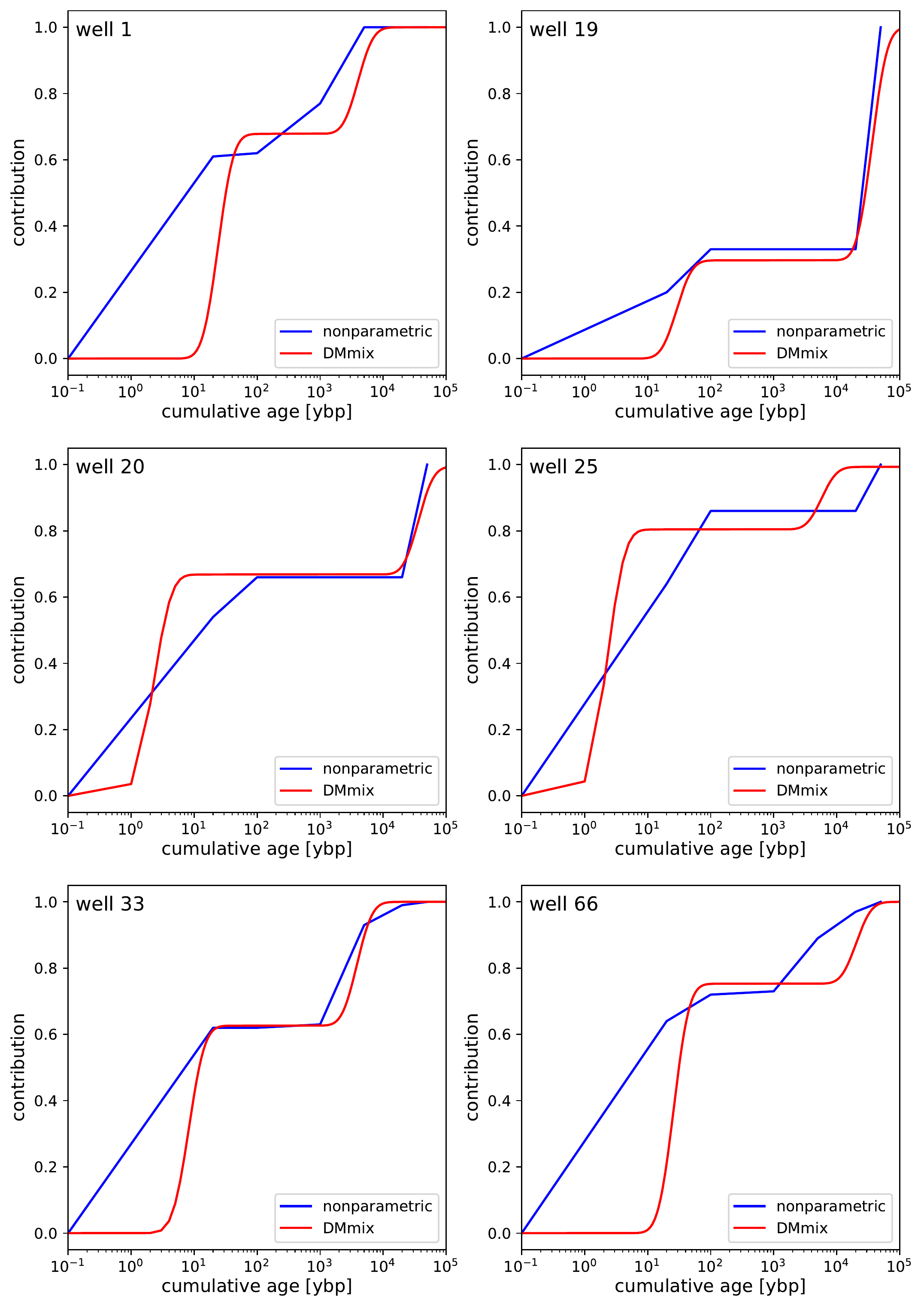}
    \caption{Cumulative age distributions for the parameter estimates of both models and all wells, plotted on a logarithmic age scale. The DMmix model exhibits a more abrupt behaviour, i.e. steeper slopes and longer horizontal stretches than the nonparametric model. }
    \label{fig:sup_age_distrib}
\end{figure}

\newpage

\noindent\textbf{Figure S4 (Constraining the flow regime for DMmix)}
\begin{figure}[H]
\setcounter{figure}{3}
    \centering
    \includegraphics[width=\linewidth]{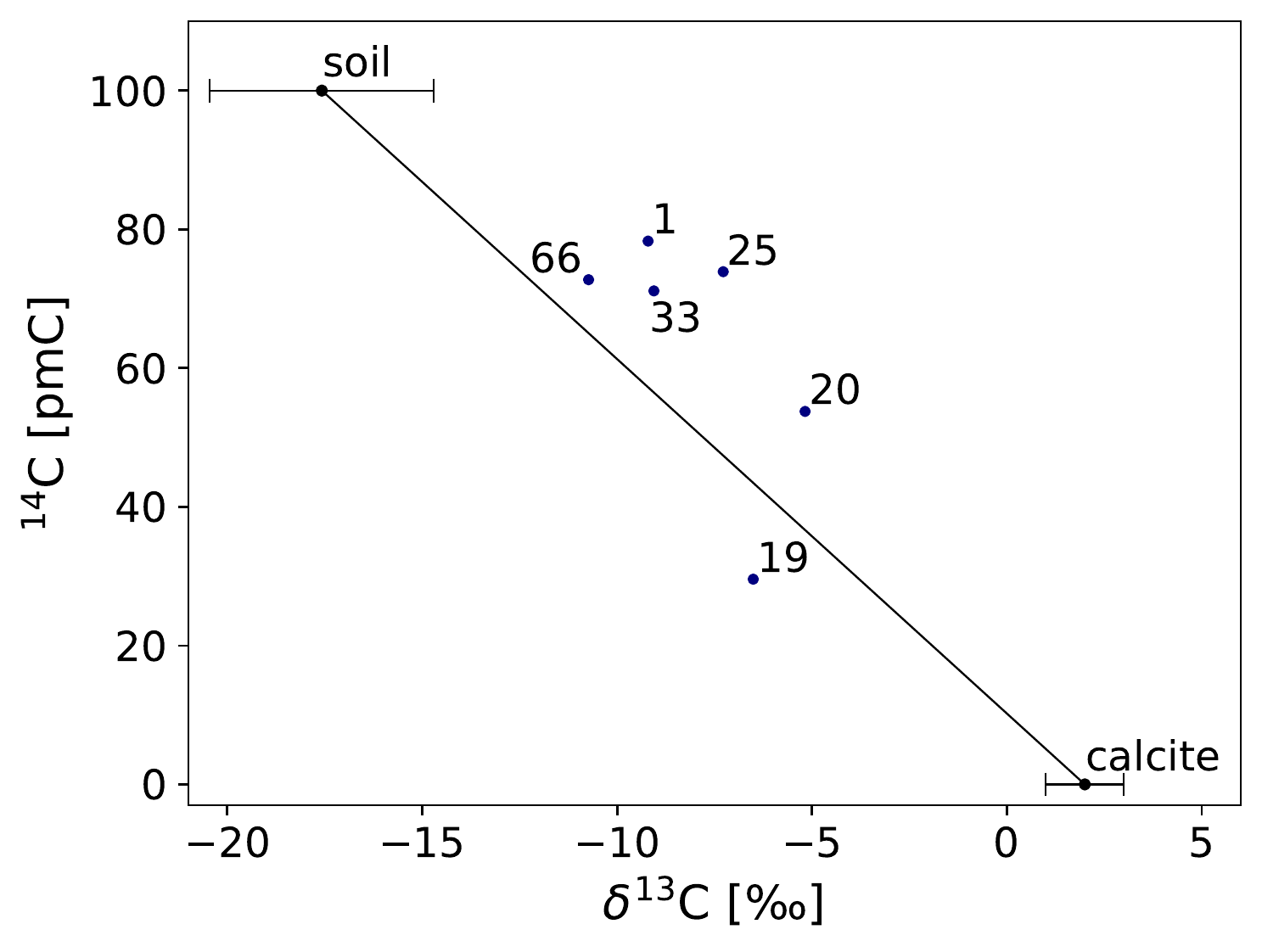}
    \caption{Modelling of the \textsuperscript{39}Ar, \textsuperscript{14}C and CFC-11 tracer data using the DMmix model (a Dispersion Model for a young (1) and an old (2) water mass) with different Péclet numbers. Wells 1, 20, 33 and 66 show uniform likelihoods whereas wells 19 and 25 exhibit opposing trends. As no conclusion can be drawn from this analysis, $Pe = 10$ is assumed for both water masses.}
    \label{sup:fig_peclet}
\end{figure}

\newpage
\noindent\textbf{Figure S5 (Leave-One-Out Likelihood of an example well)}
\begin{figure}[H]
\setcounter{figure}{4}
    \centering
    \includegraphics[width=\linewidth]{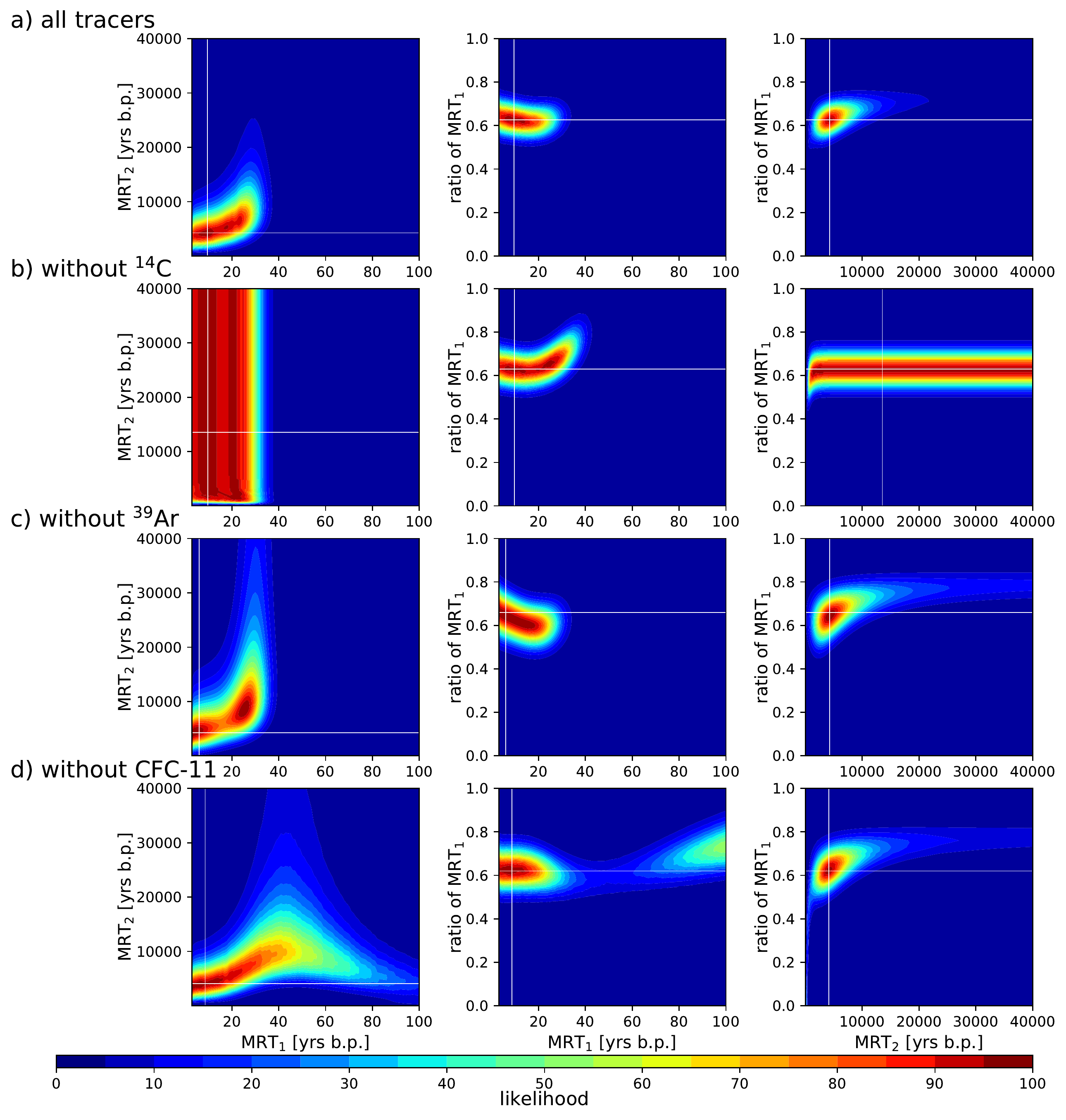}
    \caption{Likelihood score of well 33 in parameter space, a) with all tracers ($^{14}$C, $^{39}$Ar, CFC-11) b) without $^{14}$C, c) without $^{39}$Ar d) without CFC-11. Larger areas of high likelihood indicate high uncertainties of the parameter estimates. $^{14}$C is the main tracer constraining MRT$_2$, $^{39}$Ar affects all parameters slightly and CFC-11 predominantly constrains MRT$_1$.}
    \label{fig:sup_lik_loo}
\end{figure}